\documentclass{imammb}
\jno{dqnxxx}
\usepackage{natbib}
\setcitestyle{authoryear,open={(},close={)}}
\usepackage{graphicx}
\usepackage{amsmath,amsthm,bm,mathrsfs}
\usepackage{amssymb}
\usepackage{subcaption, floatrow}
\usepackage[backref=true,pagebackref=true,hyperindex=true,breaklinks=true,colorlinks=true,urlcolor=green,bookmarks=true,bookmarksopen=false,citecolor=blue,linkcolor=blue]{hyperref}
\usepackage{doi}
\usepackage[percent]{overpic}
\usepackage{xcolor}

\newcommand{\hide}[1]{}
\newcommand{\red}[1]{{{#1}}}
\newcommand{\redm}[1]{{{#1}}}

\DeclareMathOperator{\A}{\mathcal{A}}
\DeclareMathOperator{\B}{\mathcal{B}}
\DeclareMathOperator{\diag}{diag}
\DeclareMathOperator{\sgn}{sgn}
\newcommand{\Vm}{V_\text{m}}
\newcommand{\Vf}{V_\text{f}}
\newcommand{\Va}{V_\text{m}^\alpha}
\newcommand{\Vfa}{V_\text{f}^\alpha}
\newcommand{\Vo}{V_\text{m}^\omega}
\newcommand{\um}{\mathbf{u}_\text{m}}
\newcommand{\uf}{\mathbf{u}_\text{f}}
\newcommand{\ja}{j^\alpha}
\newcommand{\Cm}{C_\text{m}}
\newcommand{\Cf}{C_\text{f}}
\newcommand{\Gg}{G_\text{gap}}
\newcommand{\Gm}{G_\text{m}}
\newcommand{\Gf}{G_\text{f}}
\newcommand{\gNa}{g_\text{Na}}
\newcommand{\VNa}{V_\text{Na}}
\newcommand{\jac}{j^\alpha_\text{crit}}
\newcommand{\Nc}{N_\text{crit}}
\newcommand{\mV}{\mathrm{mV}} % millivolt
\newcommand{\ie}{i.e.}
\renewcommand{\d}{\mathrm{d}}
\newcommand{\+}[3]{\def#1{{#2}}}
\+{\Eh}{\E_h}{h-gate switch voltage}
\+{\Em}{\E_m}{m-gate switch voltage}
\+{\E}{E}{transmembrane voltage}
\+{\F}{F}{an arbirary function}
\+{\GG}{G}{Inverse of $\FF$}
\+{\N}{N}{number of abstract dyn eqns}
\+{\eps}{\epsilon}{the small parameter of the main embedding}
\+{\h}{h}{fast inward current inactivation gate}
\+{\jbar}{\Qstat{j}}{fast inward current slow inactivation gate quasistat}
\+{\l}{l}{summation index in various places}
\+{\m}{m}{fast inward current activation gate}
\+{\taul}{\tau_{\l}}{relaxation timescale of $\l$-th var}
\+{\t}{t}{original (slow) time}
\+{\w}{w}{abstract dynamic variable; aux var \eq{change-to-bessel}}

\numberwithin{equation}{section}

\allowdisplaybreaks
\begin{document}

%%%%%%%%%%%%%%%%%
\title{Action potential propagation and block in a model of atrial tissue
  with myocyte-fibroblast coupling}
\author{{\sc Peter Mortensen}\\[2pt]
School of Mathematics \& Statistics, University of Glasgow, Glasgow G12 8QQ\\
Institute of Cardiovascular \& Medical Sciences, University of Glasgow, Glasgow G12 8TA\\[6pt]  
{\sc Hao Gao}\\[2pt]
School of Mathematics \& Statistics, University of Glasgow, Glasgow
G12 8QQ\\[6pt]
{\sc Godfrey Smith}\\[2pt]
Institute of Cardiovascular \& Medical Sciences, University of Glasgow, Glasgow G12 8TA\\[6pt]
{\sc and}\\[6pt]
{\sc Radostin D.~Simitev$^\ast$}\\[2pt]
School of Mathematics \& Statistics, University of Glasgow, Glasgow
G12 8QQ\\
$^\ast${\rm Corresponding author. Email:
  \href{mailto:Radostin.Simitev@glasgow.ac.uk}{Radostin.Simitev@glasgow.ac.uk}}\\[8pt]
{\rm [Accepted on 2020-12-08]}\\[7pt]}
\pagestyle{headings}
\maketitle
\markboth{\rm MORTENSEN, GAO, SMITH \& SIMITEV}{\rm PROPAGATION \& BLOCK IN A MYOCYTE-FIBROBLAST MODEL OF ATRIAL TISSUE}
%\markboth{\rm MORTENSEN, GAO, SMITH \& SIMITEV}{}

%%%%%%%%%%%%%%%%%abstract style
%Two grouping braces are necessary in abstract environment
%first argument contains abstract text; second argument contains keywords
%text

\begin{abstract}
{
The electrical coupling between myocytes and fibroblasts and the
spacial distribution of fibroblasts within myocardial tissues are
significant factors in triggering and sustaining cardiac arrhythmias
but their roles are poorly understood.
This article describes both direct numerical simulations and an
asymptotic theory of propagation and block of electrical excitation in a
model of atrial tissue with myocyte-fibroblast coupling.
In particular, three idealised fibroblast distributions are
introduced: uniform distribution, fibroblast barrier and myocyte
strait, all believed to be constituent blocks of realistic fibroblast
distributions. Primary action potential biomarkers including
conduction velocity, peak potential and triangulation index are
estimated from direct  simulations in all cases. 
Propagation block is found to occur at certain critical values of
the parameters defining each idealised fibroblast distribution and
these critical values are accurately determined.  
An asymptotic theory proposed earlier is extended and applied to the
case of a uniform fibroblast distribution. Biomarker
values are obtained from hybrid analytical-numerical solutions
of coupled fast-time and slow-time periodic boundary value problems
and compare well to direct numerical simulations. The
boundary of absolute refractoriness is determined solely by the 
fast-time problem and is found to depend on the 
values of the myocyte potential and on the slow inactivation variable of the sodium
current ahead of the propagating pulse. In turn, these quantities are
estimated from the slow-time problem using a regular perturbation
expansion to find the steady state of the coupled myocyte-fibroblast
kinetics. The asymptotic theory gives a simple analytical expression
that captures with remarkable accuracy the
block of propagation in the presence of fibroblasts. 
}{Electrophysiology, Myocyte-Fibroblast Coupling, 
  Refractoriness, Asymptotic Approximation}
\end{abstract}
%%%%%%%%%%%%%%%%%%%%%%%%%%%%%%%

\section{Introduction}

%% Topic, Motivation, Context, Physiological background
% Arrythmias
\looseness=-1
The cardiac rhythm is controlled by electrical signals. 
The disruption of cardiac electrical signalling results in
abnormalities causing the heart to beat too slowly, too quickly, or
irregularly. These abnormalities are known as arrhythmias and their
effects may range from minor discomfort to sudden death
\citep{Huikuri2001,Qu2015}.  
% Fibroblasts
In addition to cardiac myocytes, the cells that actively generate and
transmit electrical signals in the heart,  other
non-myocyte cells play a significant role in the development of
arrhythmias and electrical dysfunction.  The most abundant type of
non-myocyte cells in the heart are the cardiac fibroblasts -- a
heterogeneous group of cells whose phenotype and main functions
vary in response to the pathological conditions of the heart
\citep{Brown2005}. 
% Fibroblast functions
In normal myocardium the fibroblasts are primarily responsible for
homeostatic maintenance of extracellular collagenous matrix. 
% phenotype
In injured myocardium the fibroblasts, activated by cytokines,
transition into a distinct myofibroblast phenotype, proliferate and
act as the key cells in the wound healing response, in particular they
secrete a collagen-based polymer matrix to support myocytes.
% regions with increased fibroblast count
Several cardiac diseases such as myocardial infarction (MI) and
dilated cardiomyopathy are thus associated with an increased density of
fibroblasts.
% Fibroblast arrhythmogenic  effects 
The arrhythmogenic action of fibroblasts is believed to occur via 
two main mechanisms (a) production of excess collagen that
decouples cardiomyocytes and causes convoluted conduction paths
resulting in substrates prone to electrical dysfunction \citep{deJong2011}
and (b) direct electrical interaction with myocytes via gap junction
channels \citep{Chilton2007,Louault2008}. The latter mechanism is 
somewhat less studied in the literature and it is the focus of this article.

%% Prior modelling
% Need for mathematical modelling
Mathematical modelling plays an important role in understanding the
arrhythmogenic electrical interactions of cardiomyocytes and fibroblasts which
are otherwise difficult to separate and study in vivo.
% Types of models available
Two types of models have been proposed for the electrical properties
of the single fibroblast -- passive and active. The active models
\citep{Jacquemet2007,Maccannel2007} incorporate the discovery that
cardiac fibroblasts express voltage-gated K+ currents
\citep{Chilton2005}, while the passive models \citep{Nayak2013}
regard fibroblasts as inert electrical loads. While 
most models describe ventricular 
fibroblasts, atrial versions \citep{Burstein2008} and myofibroblast phenotype
versions \citep{Chatelier2012,Koivumki2014} are also available.
For upscaling to tissue level either a cell-insertion approach or
a cell-attachment approach is commonly adopted
\citep{Xie2009}. Single-cell and tissue models have been used to study
a variety of  
effects of myocyte-fibroblast electrical coupling including changes in
action potential morphology \citep{Maccannel2007}, excitability  and
conduction velocity \citep{2006Miragoli,Xie2009}, spontaneous
self-excitation \citep{Miragoli2007,Greisas2012}, propensity to 
early after-depolarisation and cardiac alternans \citep{Nguyen2011},
and vulnerability to reentry \citep{Majumder2012,Gomez2014,Gomez2014a}. 
Modelling tends to veer towards using realistic patient-specific 3D
computational models \citep{McDowell2011}, an approach that, while
clinically most relevant, due to its complexity is not well-suited to
disentangle and explain causal effects and possible mechanisms of 
arrhythmogenesis.

%% Goals of paper
% Conduction velocity, refractoriness and restitution
Conduction disturbances, spatially non-uniform conduction and conduction block
are thought to be key elements in the initiation and maintenance of
one of the two main types of arrhythmias, the reentrant arrhythmias,
of which tachycardia, atrial and ventricular fibrillation are prominent
examples \citep{Antzelevitch2011,Qu2015}.
Reentry around anatomic or functional obstacles is determined by
both conduction velocity and refractoriness. A reentrant arrhythmia
occurs when the length of a circuit exceeds the wavelength of the
circulating excitation given by the product of its conduction velocity
and its refractory period.
% Aim
Thus, to understand reentrant arrhythmogenesis in post-MI fibrous
substrates we aim to model conduction velocity and refractoriness in
tissues composed of coupled myocytes and  fibroblasts.
% Related work that we wish to extend
While conduction velocity or equivalently activation times are
routinely measured experimentally e.g. \citep{Dietrichs2020,Erem2011} and
computed from direct numerical simulations e.g.~\citep{Xie2009} there are few
attempts for rigorous theoretical analysis, see discussions in
\citep{Tyson-Keener-1988,Meron1992}. These attempts have been based 
almost universally  on conceptual models of FitzHugh-Nagumo type
\citep{FitzHugh-1961,Nagumo-etal-1962} that are incapable of reproducing 
slow repolarisation, slow sub-threshold response, fast accommodation
and other features of cardiac excitability crucial for understanding
and controlling arrhythmogenesis \citep{Biktashev2002,Biktashev2008}. 
To address these deficiencies we developed earlier an asymptotic
procedure for analysis of detailed cardiac ion current models \citep{Biktasheva2006,Biktashev2008}.
The key step consists of embedding asymptotically small parameters
in the detailed models considered based on (a) a formal analysis of the time-scales of
evolution of dynamic variables,  (b) the largeness of the maximal
value of the sodium current $I_\mathrm{Na}$ compared to other currents
and (c) the quasi-stationary permeability of the $I_\mathrm{Na}$ ionic
gates in certain potential ranges.  
The asymptotic embedding procedure reveals that, unlike systems  of
FitzHugh-Nagumo type, cardiac models have non-Tikhonov asymptotic
structure \citep{Tikhonov-1952,Pontryagin-1957} including qualitatively
new features of topological nature \citep{Simitev2011}.
Asymptotic model reduction based on such embedding is capable of
reproducing essential spatiotemporal phenomena of cardiac electrical
excitation as demonstrated in \citep{Simitev2011} using several
different cardiac ionic models. In \citep{2006Simitev}, a simplified 
version of the procedure was used to achieve numerically accurate
prediction of the front propagation velocity (within 16\%) and its
profile (within 0.7 mV) for a realistic model of human atrial tissue
\citep{CRN98}. The asymptotic reduction was sufficiently
simple to allow the derivation of an analytical condition for
propagation block in a re-entrant wave which was in an excellent
agreement with results of direct numerical simulations of the
realistic atrial ionic model.

% Specific goals of the article
In this work, we wish to adapt and extend the analysis of
\citep{2006Simitev} in order to study conduction velocity and
conduction block in tissues composed of coupled myocytes and
fibroblasts. In particular, we wish to demonstrate that the asymptotic
theory discussed above will also apply with little modification in
the case of myocyte-fibroblast coupling.
However, we expect that conduction velocity will depend on the ratio
of fibroblasts to myocytes in the tissue via changes that coupling
elicits to the resting potential and to the action potential shape
\citep{Jacquemet2008,Weiss2009}. Further to this, we wish to use the 
asymptotic theory to explain the critical ratio of fibroblasts to
myocytes at which a homogeneous and isotropic patch of fibrous tissue
will block conduction and act as an anatomical obstacle. In reality,
fibrous tissue is, of course, highly inhomogeneous and anisotropic
\citep{Weiss2014,Wald2018}. Arrhythmogenic risk predominantly arises in the border zone
of a fibrotic post myocardial infarction zone where the boundaries
between necrotic and morphologically viable myocardium are nonlinear and
characterized by irregular edges wherein islands or bundles of
myocytes are interdigitated by scar tissue 
\citep{Arbustini2018}. While details of the scar morphology depend  
on a multitude of factors and very patient-specific, it seems possible
to identify a small number of commonly occurring, simple fibrotic
inhomogeneities and investigate their transmission capacity. We focus
in particular, on two seemingly typical fibrotic 
inhomogeneities that we will refer to as ``straits'' and ``barriers''. 
Straits can be described as channels collinear with the direction of
electrical conduction that have small fibroblast-to-myocyte ratio and
are bordered on both sides by regions with large fibroblast-to-myocyte
ratio. Barriers can be described as strips perpendicular to the
direction of conduction that have large fibroblast-to-myocyte ratio
and separate regions of small fibroblast-to-myocyte ratio.
We suggest that other more complicated fibroblast distributions seen
in histology can be understood on the basis of these and other basic inhomogeneities. 

% Structure and exposition
The article is structured as follows.
In Section \ref{sec:2}, we formulate our mathematical model of fibrous
atrial tissue including three particular versions of the monodomain
equations used, the form of myocyte-fibroblast coupling current,
parameter values, model kinetics 
and auxiliary conditions. We proceed to describe three idealised
fibroblast distribution cases that we believe are constituent blocks
of realistic fibroblast distributions. We also briefly summarize the
Strang operator-splitting method used for numerical solution.
In Section \ref{sec:3}, we present results from direct numerical
simulations of the three idealised fibroblast distributions. In
particular, we report simulated values of the primary action potential
biomarkers routinely measured in experiments on electrical excitation
in myocardial tissues e.g.~by optical mapping. We also
measure the critical values of the parameters defining the three
idealised fibroblast distributions at which action potential
propagation failure occurs. 
In Section \ref{sec:4}, we extend and apply the asymptotic theory of
\cite{Biktashev2008,Simitev2011} to the case of uniform fibroblast
distribution. Appropriate fast-time and slow-time systems are derived
and coupled asymptotically. Solutions are of these asymptotic problems
are obtained, notably analytical expressions for the resting steady
state of the coupled myocyte-fibroblast kinetics using a regular
perturbation expansion. 
We conclude with a discussion of results and questions for further work in Section \ref{sec:5}.

\section{Mathematical models and numerical methods} 
\label{sec:2}
\subsection{Mathematical model of fibrous atrial tissue}

To model the propagation of electrical excitation in two-dimensional
fibrous atrial tissue we use the monodomain equations, see
e.g.~\citep{Sundnes2006,ColliFranzone2014}, in the modified form,
\begin{subequations}
\label{e:tissue}
\begin{align}
\chi\left(C_\text{m} \frac{\partial V_{\text{m}}}{\partial t} +
I_{\text{m}}(V_{\text{m}}) +
n(\mathbf{x})G_{\text{gap}}(V_{\text{m}}-V_{\text{f}}) +
I_{\text{stim}}(\mathbf{x},t)\right) &= \nabla\cdot(\mathbf{\sigma} \nabla V_{\text{m}}) \quad \text{in} \quad \Omega \times
(0,\infty),  \label{e:Vm} \\
%I_{\text{stim}}(\mathbf{x},t)\right) &= \nabla\cdot(\mathbf{\sigma} \cdot \nabla V_{\text{m}}) \quad \text{in} \quad \Omega \times (0,\infty),  \label{e:Vm} \\
C_\text{f} \frac{\partial V_{\text{f}}}{\partial t} +
I_{\text{f}}(V_{\text{f}}) +
G_{\text{gap}}(V_{\text{f}}-V_{\text{m}}) &=   0 \quad \text{in} \quad \Omega \times (0,\infty),  \label{e:Vf} \\
\mathbf{k}\cdot\nabla V_\text{m} & = 0 \quad \text{in} \quad
\partial\Omega \times (0,\infty).  \label{e:BC}
\end{align}
\end{subequations}
The problem is posed in a rectangular region $
\Omega =[0,L_x]\times[0,L_y] \subset \mathbb{R}^2$ of sizes $L_x$ and $L_y$ in the $x$- and the
$y$-directions, respectively, with a unit normal $\mathbf{k}$ to its
boundary $\partial \Omega$. Time is denoted by $t$, and the position
vector is given by $\mathbf{x} = (x,y)$ in a Cartesian coordinate
system shown in Fig.~\ref{fig:FibrosisLabel}. Here, 
$V_\text{m}(\mathbf{x},t)$ and $V_\text{f}(\mathbf{x},t)$ denote 
myocyte and fibroblast transmembrane potentials and 
$C_\text{m}$  and $C_\text{f}$ are myocyte and fibroblast membrane
capacitances, respectively. The constant $\chi$ denotes cell 
surface-to-volume ratio, and $\mathbf{\sigma}$ is a constant,
second-order conductivity tensor, with an anisotropy ratio of
approximately 15:2. The terms $I_\text{m}(V_\text{m})$ and $I_\text{f}(V_\text{f})$
represent the sum of voltage-dependent ionic currents across myocyte and 
fibroblast membranes as defined in the realistic human atrial cell
model of \cite{CRN98} and in the active fibroblast model of
\cite{Morgan2016}, respectively.  
The stimulus current $I_\text{stim}(\mathbf{x},t)$ has a
profile of a periodic train of rectangular impulses with amplitude
$I_s$, extent $x_s$, duration $t_s$, and period (basic cycle length) $B$,
\begin{gather}
I_\text{stim}(x,y,t)=I_s\, H(x_s-x) \left[1+ \sgn \left(\sin\frac{\pi t}{B}\right) \sgn \left(\sin\frac{\pi(t-B-2 t_s)}{B}\right)\right]
\label{eq:stim}
\end{gather}
where $H(x)$ is the Heaviside step function. Default
parameter values are listed in Table \ref{t:pars} and used throughout
unless specified otherwise. 
Resting values of the dynamical variables are used as initial
conditions as discussed further in Section \ref{sec:3}.
\begin{table}[t!]
{  \small
  \begin{tabular}{ llclll  }
    \hline
    Symbol   &	Parameter   &Value& Unit & Source Ref.\\
    \hline
    $L_x$           & domain length & 50 & mm & --\\
    $L_y$           & domain width  & 10 & mm & --\\
    $C_\text{m}$    & myocyte capacitance              &  100  &   pFmm$^{-2}$  & \citep{CRN98} \\
    $C_\text{f}$    & fibroblast capacitance           &  6.3  &   pFmm$^{-2}$ &  \citep{Maccannel2007}\\
    $G_\text{gap}$  & intercell conductance            &  0.5  &   nS    &  \citep{Morgan2016} \\
    $\chi$          & surface-to-volume ratio &  140  &   mm$^{-1}$ &    \citep{Niederer2011}\\
    $\sigma(\mathbf{x})$  & conductivity tensor & $ \begin{pmatrix} 
      0.1334 & 0 \\
      0 & 0.0176 
    \end{pmatrix}$  &   Sm$^{-1}$   &  \citep{Niederer2011} \\
    $I_s$ & stimulus current amplitude &-2000 &pA& \citep{CRN98} \\
    $x_s$ & stimulus extent &1 &mm& \citep{CRN98} \\
    $t_s$ & stimulus duration & 2 &ms& \citep{CRN98} \\
    $B$   & basic cycle length & variable &  ms &--\\
    $I_\text{m}(V_\text{m})$ & myocyte current kinetics & variable &pA& \citep{CRN98}\\
    $I_\text{f}(V_\text{f})$ & fibroblast current kinetics & variable &pA& \citep{Morgan2016}\\
    \hline
  \end{tabular}}
  \caption{Default values of model parameters and kinetics used in equations \eqref{e:tissue}.}	
\label{t:pars}
\end{table}

% Comments
%% CRN model
The realistic human atrial cell model of \citet{CRN98} used here
includes all major transmembrane currents responsible for AP
generation such as $I_\text{Na}$, $I_\text{Kur}$, $I_\text{to}$,
$I_\text{Kr}$, $I_\text{Ks}$, $I_\text{CaL}$, $I_\text{K1}$,
$I_\text{NaK}$, and $I_{NCX}$. The model also incorporates a description
of intracellular Ca$^{2+}$ handling  that accounts for the uptake and
release currents $J_\text{up}$ (SERCA) and $J_\text{rel}$ (RyR) of the
sarcoplasmic reticulum as well as K$^{+}$, Na$^{+}$ and Ca$^{2+}$
homeostasis regulating intracellular ionic concentrations. We have
chosen to use this model as it gives the opportunity to validate our
numerical codes against data available in the literature, and to use
asymptotic results already  obtained in this case by \cite{2006Simitev}. 
%% homogeneous conductivity
Constant values for the components of the conductivity tensor $\sigma$
are assumed in equation \eqref{e:Vm} since we wish to focus the
attention on the arrhythmogenic role of direct electrical interaction
between myocytes and fibroblasts rather than on the better studied
effects of excess collagen density in fibrotic regions. 

% Myocyte-fibroblast coupling
In equations \eqref{e:tissue} a fixed number of identical fibroblasts, $n(x,y)$,
are connected in parallel to each myocyte via an inter-cell conductance $G_\text{gap}$. 
This arrangement represents the so called ``single-sided'' fibroblast-myocyte
connection proposed by \cite{Kohl2007} where fibroblasts electrically
couple to one (or more) myocytes that are themselves electrically
well-connected to each other while there are no fibroblast-fibroblast connections. 
A similar arrangement was referred to as ``fibroblast-attachment'' in \citep{Xie2009}.

\subsection{Idealised fibroblast distributions}
\begin{figure*}[!t]
\includegraphics[width=\textwidth]{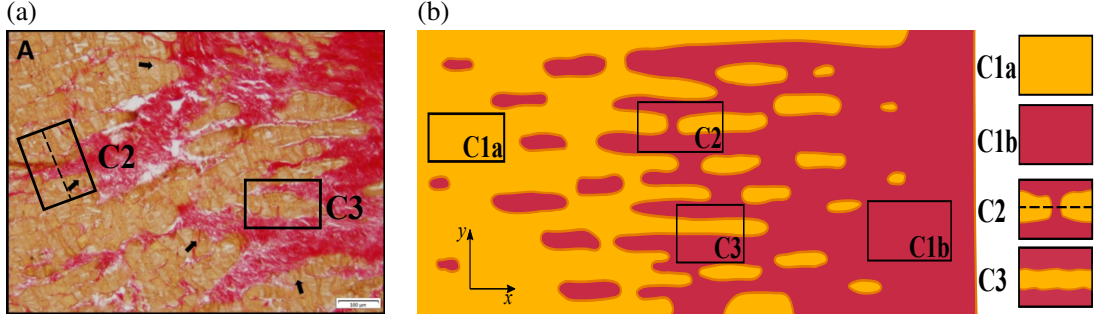}\\[2mm]
% \begin{center}
% \begin{tabular}{ll}
% (a) & (b) \\
% {\includegraphics[height=0.25\linewidth]{Fig01a.eps}} &
% %\hfill
% \includegraphics[height=0.25\linewidth,width=0.6\linewidth]{Fig01b.eps} \\
% \end{tabular}
% \end{center}
\caption{(a) An example of fibrosis in a specimen of cardiac tissue
with collagen stained  in red and intact myocardium shown in yellow from
\cite{Wald2018}.
(b) A  schematic illustration of a border zone between intact myocardium
(to the left) and fibrosis (to the right) where the darker shaded
areas indicate increased  fibroblast density. The rectangular regions denoted
    in (b) correspond to the specific forms of the fibroblast
    distribution function $n(x,y)$ considered, see text.
}
\label{fig:FibrosisLabel}
\end{figure*}

The function $n(x,y)$ can be used to prescribe realistic non-uniform fibroblast
distributions of variable severity such as reported in
\citep{Weiss2014,Wald2018,Arbustini2018}. However, here we restrict
the attention to three simple cases illustrated in Fig.~\ref{fig:FibrosisLabel}(b) that are perhaps characteristic of the distributions arising in border zones between
intact myocardium and mature scars, an example shown in Fig.~\ref{fig:FibrosisLabel}(a).
In particular, we consider the following profiles of $n(x,y)$. 
\begin{enumerate}
\item[C1.]  Uniform fibroblast distribution prescribed by
\begin{subequations}
  \label{nx}
\begin{gather}
  n(x,y)=N,
\label{n1}
\end{gather}
where $N$ is a constant over the entire domain. This case is
schematically illustrated in boxes C1a and C1b of Fig.~\ref{fig:FibrosisLabel}(b) with $N=0$ describing intact myocardium with no fibroblasts and $N>0$
indicating $N$ fibroblasts attached to each myocyte, respectively.

\looseness=-1
\item[C2.] ``Fibroblast barrier'' distribution prescribed by
\begin{gather}
n(x,y)=N\, H\big((L_x/2+\Delta x)-x\big) H\big(x-(L_x/2-\Delta x)\big),
  \label{n2}
\end{gather}
where intact myocardium with no attached fibroblasts is separated by a
rectangular region of constant width $\Delta x$ extending
uniformly in the $y$-direction where $N$ fibroblasts are attached to each 
myocyte. This case is a simplification of the situation schematically
illustrated in box C2 of Fig.~\ref{fig:FibrosisLabel}(b).

\item[C3.] ``Myocyte strait'' distribution prescribed by
\begin{gather}
n(x,y)=N\, \big( H\big((L_y/2-\Delta y)-y\big) + H\big(y-(L_y/2+\Delta y)\big)\big),
\label{n3}
\end{gather}
\end{subequations}
where intact myocardium, in the shape of a rectangular region of
constant width $\Delta y$, extends uniformly in the $x$-direction and
channelled by fibrotic regions on both sides where $N$ fibroblasts are
attached to each myocyte. This case is schematically illustrated in
box C3 of Fig.~\ref{fig:FibrosisLabel}(b).
\end{enumerate}
\vspace{-3mm}

\subsection{Numerical methods}
In order to obtain numerical solutions of the fibroblast-myocyte monodomain
equations we write system \eqref{e:tissue} in the form
\begin{gather}
\frac{\partial}{\partial t}
\begin{bmatrix}
V_\text{m}\\
V_\text{f}
\end{bmatrix}
=
\A \begin{bmatrix}
V_\text{m}\\
V_\text{f}
\end{bmatrix}
+
\B\begin{bmatrix}
V_\text{m}\\
V_\text{f}
\end{bmatrix},
\label{eq:split}
\end{gather}
where $\A$ and $\B$ are nonlinear differential operators defined by
\begin{subequations}
\label{e:opps}
\begin{gather}
\A\begin{bmatrix}
V_\text{m}\\
V_\text{f}
\end{bmatrix}
\equiv -\diag\left(\frac{1}{C_\text{m}}\Big(I_\text{m}(V_\text{m}) + n(\mathbf{x})\, G_\text{gap}(V_\text{m} - V_\text{f}) + I_\text{stim}\Big),\; \frac{1}{C_\text{f}}\Big(I_\text{f}(V_\text{f}) + G_\text{gap}(V_\text{f} - V_\text{m})\Big)\right),
\label{eq:opA}\\
\B \begin{bmatrix}
V_\text{m}\\
V_\text{f}
\end{bmatrix}
\equiv \diag\left(\frac{1}{\chi C_\text{m}}\nabla \cdot (\mathbf{\sigma}\cdot\nabla V_\text{m}), \;\;0\right).
\label{eq:opB}
\end{gather}
\end{subequations}
Following \cite{1999QuGarf}, we then apply the classical operator
splitting method of \cite{Strang1968} and numerically approximate 
the solution vector $[V_\text{m},V_\text{f}]^T_{k\Delta t}$ after $k$ time steps of
length $\Delta t$ by the following second-order accurate in
time, formal $\theta$-scheme with $\theta=1/2$
\begin{gather}
\begin{bmatrix}
V_\text{m}\\
V_\text{f}
\end{bmatrix}_{k\Delta t} =
\bigg(e^{(1-\theta)\Delta t \A}\;\; e^{\Delta t \B}\;\; e^{\theta\Delta t \A}\bigg)^k
\begin{bmatrix}
V_\text{m}\\
V_\text{f}
\end{bmatrix}_{0} + O(\Delta t^2), \qquad \qquad k=0,1,\dots,
\label{eq:strang}
\end{gather}
where $[V_\text{m},V_\text{f}]^T_0$ are specified initial conditions and
$e^X \equiv \sum_{m=0}^\infty X^m/m!$ is an operator exponential.
%and $\theta=1/2$.
\begin{figure*}[!t]
\includegraphics[width=\textwidth]{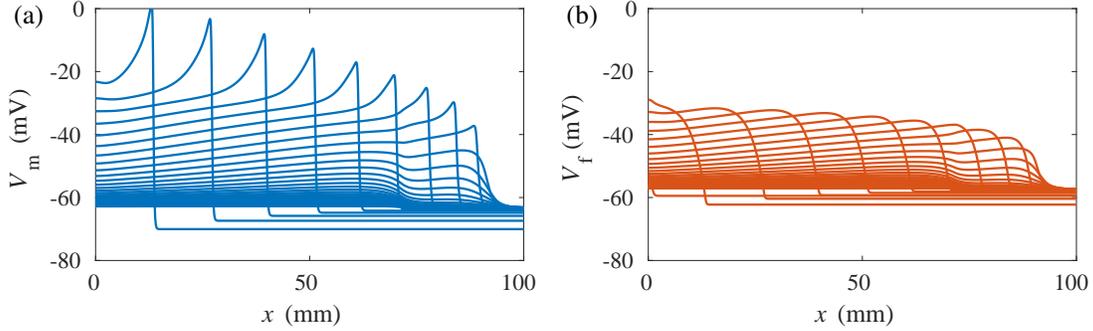}
% \setlength{\tabcolsep}{0pt}
% \begin{tabular}{l@{\extracolsep{-3mm}}l}
% \begin{overpic}[width=0.52\textwidth,tics=10]{Fig02a}
% \put (2,53) {{(a)}}
% \end{overpic}&
% \begin{overpic}[width=0.52\textwidth,tics=10]{Fig02b}
% \put (2,53) {{(b)}}
% \end{overpic}
% \end{tabular}
\caption{
\looseness=-1
  Myocyte and fibroblast action potentials $V_\text{m}$ in (a) and
  $V_\text{f}$ in (b), respectively, propagating in the $x$-direction in
  tissue with uniform fibroblast distribution \eqref{n1}. Propagation
  is illustrated by potential profiles plotted at 25\,ms time intervals
  apart. Fibroblast count $N$=12, other parameter values are listed in Table \ref{t:pars} and $B\to\infty$.}  
\label{fig:FibN12}
\end{figure*}

The practical implementation of the scheme consists of splitting the
monodomain problem \eqref{e:tissue} into three Cauchy problems coupled
to each other via their initial conditions as formally represented 
by the three nested operator exponentials in equation \eqref{eq:strang}.
The problems represented by the exponentials of $\A$ consist
of a spatially-decoupled nonlinear system of stiff ordinary
differential equations and are solved using an fourth-order implicit
Runge-Kutta method.
The problem represented by the exponential of $\B$ is a linear
diffusion equation which is spatially discretised using a low-order
finite element scheme and implemented in 
the open-source parallel C++ finite element library \texttt{libMesh}
\citep{libMeshPaper} following the implementation reported by
\cite{rossi2017incorporating}. The numerical simulation code has been
validated in \cite{Mortensen2018} against the benchmark paper of
\cite{Niederer2011}. Due to the stiffness of the ionic current system, a
high-resolution spatial mesh with typical size 0.1\,mm and a time step
of 0.005\,ms is required to resolve the upstrokes of propagating
action potentials. 

\section{Direct numerical simulations of propagation in fibrous tissue}
\label{sec:3}
\looseness=-1
In this section, we present results from direct numerical simulations
of equations \eqref{e:tissue} for the three choices of fibroblast
distribution \eqref{nx} introduced in Section \ref{sec:2}. In each of
these cases we provide numerical values of selected biomarkers
typically used to characterise  propagation in experimental measurements. In particular,
we report values  of conduction velocity, peak potential, 
peak intracellular calcium transient, APD$_\text{90}$ duration and triangulation
index as functions of the 
number of fibroblasts per myocyte. Action 
potential excitation and conduction are inhibited as fibroblast count
increases, and we proceed to estimate the critical parameters of the
fibroblast distributions \eqref{nx} at which propagation block occurs.     
\begin{figure*}[!t]
  \includegraphics[width=\textwidth]{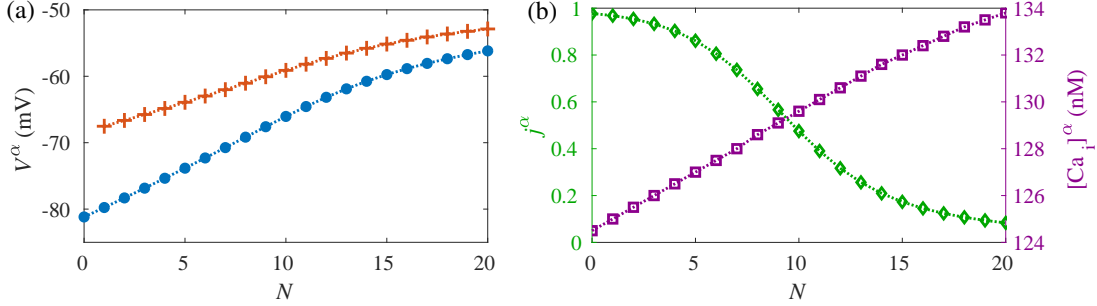}
% \setlength{\tabcolsep}{0pt}
% \begin{tabular}{l@{\extracolsep{-3mm}}l}
% \begin{overpic}[width=0.5\textwidth,tics=10]{Fig03a.eps}
% \put (0,54) {{(a)}}
% \end{overpic}&
% \begin{overpic}[width=0.55\textwidth,tics=10]{Fig03b.eps}
% \put (2,49) {{(b)}}
% \end{overpic}\\
% \end{tabular}
\caption{
(a) Resting values of the myocyte potential $V^\alpha_\text{m}$ (blue full
circles) and of the fibroblast potential $V^\alpha_\text{f}$, (red
crosses), (b) resting values of the myocyte gating variable $j^\alpha$
(green diamonds, left ordinate scale) and of the internal myocyte
calcium concentration (violet squares, right ordinate scale) all as functions of the
fibroblast count $N$ in the case of uniform fibroblast distribution
\eqref{n1}. Other parameter values are listed in Table \ref{t:pars} and $I_s=0$. }
\label{fig:RestingPotentialPlot}
\end{figure*}

\subsection{The case of uniform fibroblast distribution (C1)}
\label{UniformC1}
\looseness=-1
The basic effects of myocyte-fibroblast electrical coupling on
conduction are best understood in the simple case C1 of uniform
fibroblast distribution \eqref{n1} which we proceed to discuss here.
Fig.~\ref{fig:FibN12} shows a numerical \red{thought experiment}
that illustrates \red{visually many of the phenomena analysed
further in the text and their underlying mechanisms.} Here, a tissue
model with $N=12$ fibroblasts coupled to each myocyte is considered.
Myocyte and fibroblast variables are set initially to the resting
values of uncoupled cells as defined in \cite{CRN98} and
\cite{Morgan2016}, respectively. 
A stimulation by a current injection with a single impulse
(i.e.~$B\to \infty$ in \eqref{eq:stim}) is performed. The myocyte and
fibroblast transmembrane  potentials assume  typical action potential
profiles as shown in Fig.~\ref{fig:FibN12}. As the signals propagate
in the $x$-direction and in time, their peak values $V_\text{m}^\omega$ and
$V_\text{f}^\omega$ decrease while the pre-front potentials
$V_\text{m}^\alpha$ and $V_\text{f}^\alpha$ increase. The overall shapes
of the action potentials change and most notably the steep profile of
the myocyte potential, $V_\text{m}$ is eroded and assumes a much more
diffusive profile than initially. This coincides with a slowdown of
conduction until eventually decay of the action potentials
occurs. Since conduction velocity and action potential features depend on the state
of the medium ahead of the front this behaviour in our experiment can
be explained by the process of relaxation of pre-front values of the
myocyte and fibroblast potentials, gating variables, and ionic
concentrations to their resting states. This relaxation occurs simultaneously with
the propagation of the AP into the tissue.
The resting state of the coupled myocyte-fibroblast system is different from the resting states of the
uncoupled myocytes and fibroblast cells and Fig.~\ref{fig:RestingPotentialPlot} shows the equilibrium values
$V^\alpha_\text{m}$ and $V^\alpha_\text{f}$ of myocyte and fibroblast
potentials and, as examples, also of the slow inactivation gating
variable of the myocyte sodium current $j^\alpha$ and of the resting calcium concentration,
$\text{[Ca}_\text{i}]^\alpha$, all as functions of the number of coupled 
fibroblasts $N$. For each value of $N$, the resting values are computed by
suppressing stimulation and leaving the tissues to relax for 1000 ms
at which moment the equilibrium values are recorded. The pre-front
potentials $V^\alpha_\text{m}$ and $V^\alpha_\text{f}$ as well as the
$\text{[Ca}_\text{i}]^\alpha$ increase monotonically while $j^\alpha$
decreases from their respective uncoupled values with 
the increase of the number of coupled fibroblasts $N$.
\begin{figure*}[!t]
\includegraphics[width=\textwidth]{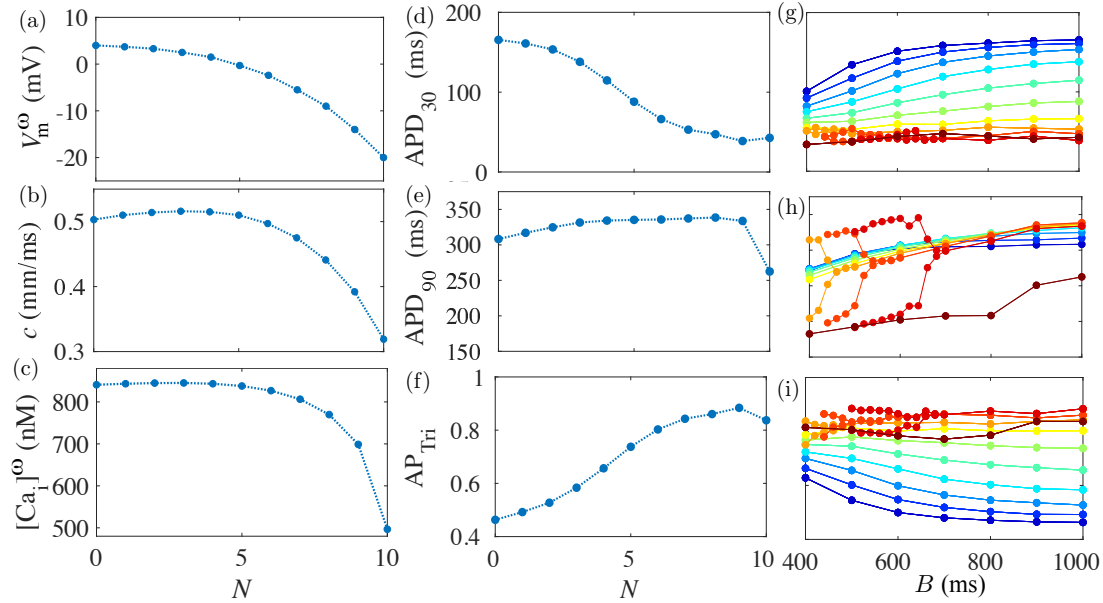}
	\caption{(a) Peak myocyte potential, $\Vo$, (b) AP wave
          speed, $c$, (c) peak internal myocyte calcium concentration,
          (d) myocyte APD$_{30}$ (e) myocyte APD$_{90}$ and (f)
          normalised triangulation index, AP$_\text{Tri}$, as
          functions of the number of coupled fibroblasts $N$ in the
          uniform case \eqref{n1} and for a single stimulus
          $B\to\infty$. (g) Myocyte APD$_{30}$ (h) myocyte APD$_{90}$
          and (i) normalised triangulation index, AP$_\text{Tri}$, 
          as functions of basic cycle length $B$. In (h-i) the
          coloured lines represent different numbers of coupled
          fibroblasts, with dark blue corresponding to $N=0$ and dark red corresponding to
          and $N=10$, respectively, with the rest of the values on           a heat map color scale.}
\label{fig:Case1Results}
\end{figure*}

Fig.~\ref{fig:Case1Results} shows the dependences of other selected
AP biomarkers commonly measured experimentally on the number of
coupled fibroblasts $N$.   
\red{The most significant result is} the existence of a critical number
of fibroblasts beyond which action potential propagation is inhibited
and the tissue relaxes to equilibrium soon after stimulation. At the
parameter values used in \red{Fig.}~\ref{fig:Case1Results} this critical number is $\Nc=10$.
At values of $N$ smaller than this critical value, normal action
potentials are established and travel in the $x$-direction with constant
wave speed, $c$, and fixed shape. These can be characterised with the
value of their peak potential, $V^\omega$, the action potential
duration, APD$_{90}$, defined as the time taken for the AP to return to 90\% repolarisation after the
initial depolarisation, and  the normalised triangulation index,
\begin{equation*}
   \text{AP}_\text{Tri} = \frac{\text{APD}_{90}-\text{APD}_{30}}{\text{APD}_{90}}.
\end{equation*}
% wave speed
Wave speed values, $c$, exhibit non-monotonic behaviour with a slight increase in the interval $N \in[0,4]$ and decrease for larger
values of $N$ until $\Nc$ is reached as shown in Fig.~\ref{fig:Case1Results}(b). This behaviour is similar to the ``biphasic'' 
behaviour reported by \citet{Miragoli2006} in experiments and by
\citet{Xie2009} in simulations of a cell-attached model. This 
non-monotonicity was suggested to occur because wave speed first increases by
the fibroblast bringing the membrane potential closer to the threshold
for sodium current activation but then decreases as the increasing
fibroblast density shift the cardiomyocyte membrane resting potential
and sodium inactivation as discussed above. This will be subject to
further theoretical modelling in Section \ref{sec:4} further
below.
% AP shape
With the increase of the number of coupled fibroblasts $N$, both the peak potential 
values $V^\omega$ (Fig.~\ref{fig:Case1Results}(a)) and the values of
APD$_{30}$ (Fig.~\ref{fig:Case1Results}(d)) decrease, while the
values of APD$_{30}$ (Fig.~\ref{fig:Case1Results}(e)) somewhat
increase, before collapse at $\Nc=10$, giving rise to increasingly triangular 
AP profile as measured by APD$_\text{Tri}$
(Fig.~\ref{fig:Case1Results}(f)).
% Calcium
Another important quantity is the internal myocyte calcium
concentration, $\text{[Ca}_\text{i}]$ which is directly linked the
magnitude of myocyte contraction, and is used to couple
electrophysiological models to models of sarcomere mechanics
\citep{Rice2008}. The peak calcium concentration shown in Fig.~\ref{fig:Case1Results}(c) stays relatively constant until after $N=6$,
after which it decreases. Without coupling the EP model to a model of
contraction, it is clear that coupling a high number of fibroblasts
will affect muscle contraction significantly.
\begin{figure}[!t]
	\centering
        \includegraphics[width=\textwidth]{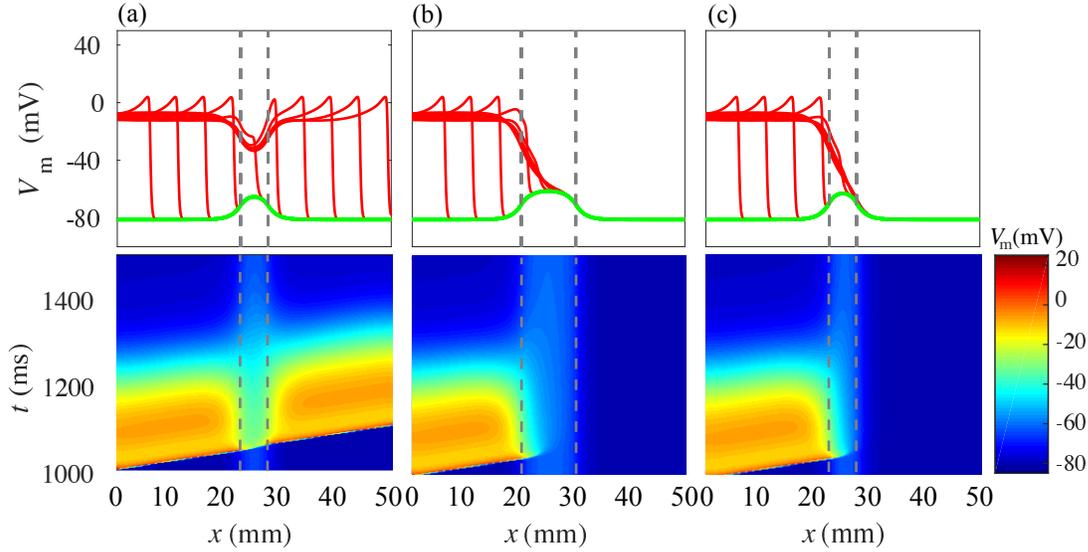}\\[2mm]
	\caption{Examples of propagation and block of action potentials
          travelling in a direction perpendicular to a fibroblast
          barrier case C2 \eqref{n2} with (a) $N=15$ and $\Delta
          x =5$ mm, (b) $N=15$ and $\Delta x=10$ mm, and (c)
          $N=18$ with $\Delta x=5$ mm, respectively.
          The top row shows cross-sections in the $x$-direction of the myocyte
          transmembrane potentials at the middle of the simulation
          domains, $L_y/2$, and at time intervals of 10\,ms after
          1000\,ms of initial relaxation, i.e. $V_\text{m}(x, y=L_y/2, t= 1000+10
          k)$, $k=0,1,2 \dots$ The resting profile at
          $t=1000$\,ms is shown by a thick green line. The bottom row
          shows density map plots of $V_\text{m}$ as a function of the
        %   horizontal 
          $x$-direction and time at $y=L_y/2$.
        The thin grey broken vertical lines show the location
          and the width $\Delta x$ of the fibroblast barriers in each
          case. }
	\label{fig:XGapComplete}\vspace*{-9pt}
\end{figure}

% BCL dependence
\looseness=-1
\red{Usually, tissue is paced periodically in vivo as well as in vitro.}
To mimic this we investigated the effects that changing the basic
cycle length $B$ (BCL) has on the action potential profile and its
propagation. Simulations were performed for $N< \Nc$ for a
physiological range of BCLs ranging from 300ms (200bpm) to 1000ms
(60bpm). To allow the stimulated APs to adjust to the BCL a tissue of
length 20mm was simulated for 6000ms, again leaving the tissue to
relax in the initial 1000ms before the first stimulation. As the BCL was
increased both the APD$_{30}$ shown in Fig.~\ref{fig:Case1Results}(g) and the APD$_{90}$
shown in Fig.~\ref{fig:Case1Results}(h) increase. However, when the
fibroblast count $N$ reaches $N=7, 8, 9$, alternans occur for
shorter BCL. 
At $N=10$  alternans do not appear, this is likely due to
the value being too close to the threshold of excitation. The
restitution curves end when the BCL becomes too short to successfully
stimulate every AP. For the smaller values of $N$, the normalised triangulation
index, AP$_\text{Tri}$ shown in Fig.~\ref{fig:Case1Results}(i) increases weakly for small
BCL, but for larger BCL the AP$_\text{Tri}$ remains relatively
constant. Action potential triangulation as an important pro-arrhythmic index \citep{Hondeghem2001}.
\begin{figure*}[!t]
  \includegraphics[width=\textwidth]{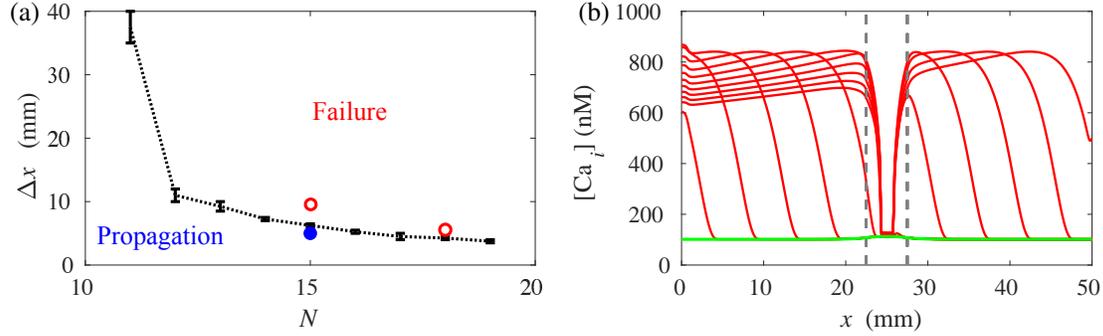}
% \setlength{\tabcolsep}{0pt}
% \begin{tabular}{l@{\extracolsep{-3mm}}l}
% \begin{overpic}[width=0.52\textwidth,tics=10]{Fig06a} 
% \put (0,54) {{(a)}}
% \end{overpic}&
% \begin{overpic}[width=0.52\textwidth,tics=10]{Fig06b} 
% \put (2,54) {{(b)}}
% \end{overpic}
% \end{tabular}
\caption{(a) The threshold curve separating the outcomes of
propagation and block of action potential in the $(N-\Delta x)$ plane
in the case of fibroblast barrier distribution \eqref{n2}. 
The error bars show the nearest pair of grid points where simulations
were run in order to construct the curve with the curve taken at 
midpoints. 
The markers show the locations of the three examples illustrated
in Fig.~\ref{fig:XGapComplete} with full blue circles indicating
successful propagation and empty red circles indicating block. (b)
Cross-sections of calcium concentration transient 
$[\text{Ca}_i](x, y=L_y/2, t= 1000+10 k)$, $k=0,1,2 \dots$
in the case $N=15$ and $\Delta x = 5$mm plotted in style similar
to the panels in the top row of Fig.~\ref{fig:XGapComplete}
Thin grey broken vertical lines show the location  and the width
$\Delta x$ of the fibroblast barrier. 
} 
\label{fig:FibNBlock}
\end{figure*}

\subsection{The case of ``fibroblast barrier'' distribution (C2)}
We now consider the idealized ``fibroblast barrier'' example of a
non-uniform fibroblast distribution defined by expression
\eqref{n2}. This case represents  healthy myocardium characterised by
small fibroblast density split in two by a fibrous 
region of high fibroblast density where $N$ fibroblasts are attached
to each myocyte. The region has a rectangular shape of constant width
$\Delta x$  extending uniformly in the $y$-direction. 
We expect that action potential propagation/failure in this case
will depend on both the width $\Delta x$ and the fibroblast count $N$ --
the two parameters needed to define this fibroblast distribution.
To illustrate this we show in Fig.~\ref{fig:XGapComplete} three
examples of action potentials propagating in the direction perpendicular
to a fibroblast barrier. Similarly to the uniform case C1, the tissue is
left to relax for 1000\,ms before stimulation commences. Panels (a)
and (b) of Fig.~\ref{fig:XGapComplete} show barrier distributions
with identical fibroblast counts $N=15$ but different widths
$\Delta x$ and illustrate that a wide barrier region can block the
propagation of an incident action potential. Panels (a) and (c)  of
Fig.~\ref{fig:XGapComplete} show barrier distributions with identical
widths $\Delta x=5$\,mm but different fibroblast counts $N$   and
illustrate that a large fibroblast count $N$  within the barrier can
also block  propagation, similarly to the uniform case of the
preceding section. These 
examples suggest that there exist critical values of the fibroblast
barrier distribution parameters $N$ and $\Delta x$ over which
block occurs. The locus of these values forms a critical curve that
serves as a threshold separating the 
outcomes of successful propagation and block in the $(N-\Delta x)$
parameter space and is shown in Fig.~\ref{fig:FibNBlock}(a).
The threshold curve plays a similar role and appears similar in shape
to ``strength-duration'' curves, familiar from experimental
electrophysiology, that serve to determine the threshold of electrical
excitation as  functions of the stimulus current amplitude and
duration. We note that as the fibroblast count $N$ approaches 10 from
above, $\Delta x$ increases asymptotically consistent with the
behaviour of the uniform case C1 discussed in the preceding section.

The threshold behaviour described above is the essential  feature of
the fibroblast barrier case C2. Because the barrier is relatively
thin its effect on the propagating action potential is only
transient if it is not blocked. In the extensive healthy regions far 
from the fibroblast barrier action potential biomarkers behave in the
same way as in the uniform case C1 described in relation to Fig.~\ref{fig:Case1Results}. 
For instance, Fig.~\ref{fig:FibNBlock}(b) shows the calcium
concentration profile propagating across a fibroblast barrier with 
$N=15$ and $\Delta x=5$\,mm. While in the barrier region the calcium
concentration is significantly less than in the ``healthy" region, it
quickly recovers on the exit of this relatively narrow strip.
\begin{figure*}[!t]
  \centering
  \includegraphics[width=\linewidth]{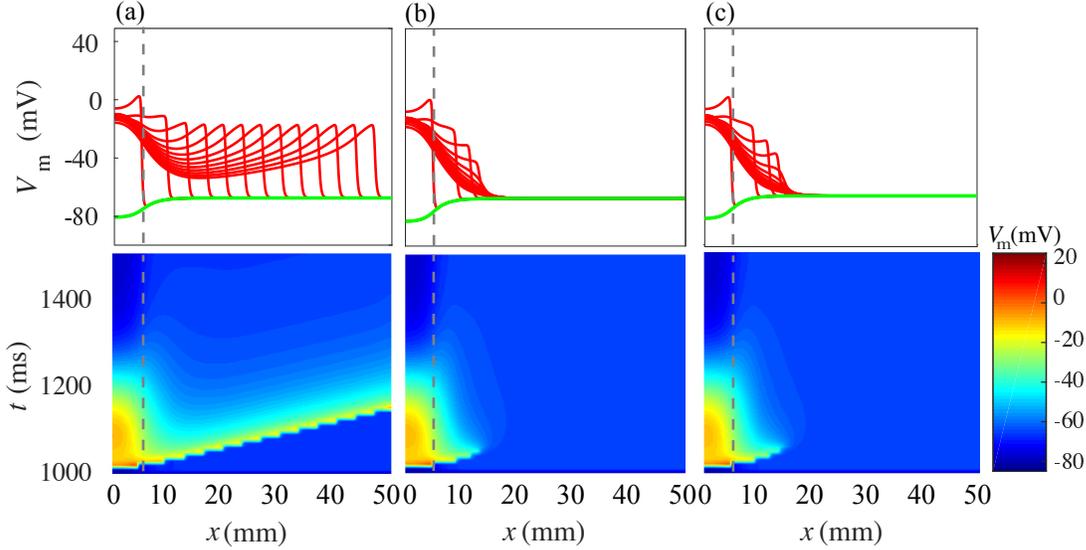} \\[2mm]
  \caption{
  Examples of propagation and block of action potentials
  travelling down myocyte straits C3 \eqref{n3} with (a) $N=15$ and
  $\Delta y=0.6$ mm, (b) $N=15$ and 
    $\Delta y = 0.4$ mm, and (c) $N=18$ and $\Delta y =0.6$ mm, respectively.
    Thin grey broken vertical lines show the left end of the
      myocyte straits and other plot elements are as described in the
    caption of Fig.~\ref{fig:XGapComplete}.}
  \label{fig:YGapComplete}\vspace*{-9pt}
\end{figure*}

\subsection{The case of ``myocyte strait'' distribution (C3)} 
\begin{figure*}[!t]
  \includegraphics[width=\textwidth]{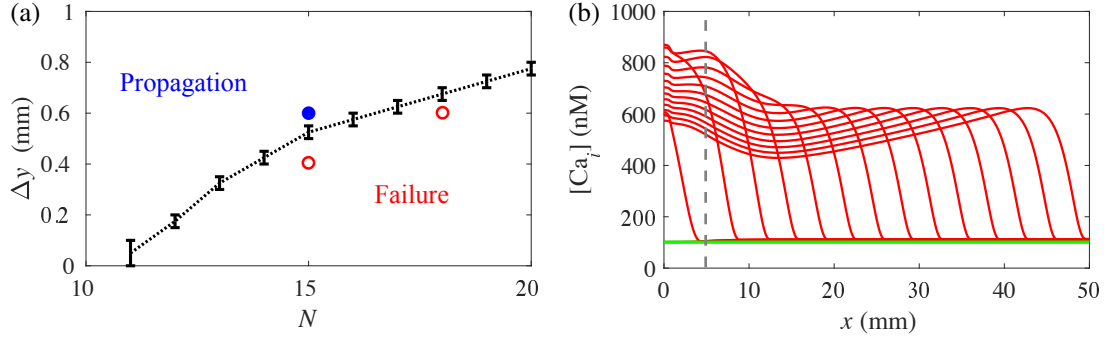}\\[3mm]
% \setlength{\tabcolsep}{0pt}
% \begin{tabular}{l@{\extracolsep{-3mm}}l}
% \begin{overpic}[width=0.52\textwidth,tics=10]{Fig08a} 
% \put (1,54) {{(a)}}
% \end{overpic}&
% \begin{overpic}[width=0.52\textwidth,tics=10]{Fig08b}
% \put (1,54) {{(b)}}
% \end{overpic}
% \end{tabular}
\caption{(a)
The threshold curve separating the outcomes of propagation and block of action potential
in the $(N - \Delta y$) plane in the case of myocyte strait
distribution \eqref{n3}. The circles represent the three examples
shown in Fig.~\ref{fig:YGapComplete}. (b) Cross-sections of
calcium concentration transient $[\text{Ca}_i]$ propagating through a myocyte
strait with $N=15$ and $\Delta y=5$ mm. Formatting conventions in both
panels are identical to these used in Fig.~\ref{fig:FibNBlock}. 
}
\label{fig:YGapLabel}\vspace*{-9pt}
\end{figure*}

\looseness=-1
Another simple idealised pattern that appears to be a
constituent of realistic non-uniform fibrosis is that of 
a relatively thin strand of viable myocytes surrounded on 
both sides by fibrous tissue.  This  distribution is defined by
expression \eqref{n3} and we will refer to it as a ``myocyte strait''
(C3). Similarly to the fibroblast barrier, a myocyte strait is
determined by two free parameters -- its width $\Delta y$ and the
fibroblast count $N$ in the adjacent fibrous regions, and similarly,
we expect that action potential propagation and failure depend on
both.
As an illustration, we show in Fig.~\ref{fig:YGapComplete} three
examples of action potentials propagating $x$-direction through 
myocyte straits. For illustrative purposes, the fibroblast distribution
used in these three cases differ from equation \eqref{n3} in that
fibroblast-free regions at $x<5$ mm have been appended in front of the
myocyte straits so that the effect of the action potentials entering the
straits can be clearly seen. As in preceding cases discussed, the
configuration is left to relax to resting state for 1000 ms, before
action potentials are stimulated.
Panel (a) of Fig.~\ref{fig:YGapComplete} shows an action potential
propagating successfully through a strait of width $\Delta y=$0.6\,mm
with 15 fibroblasts coupled to each myocyte on either side of the
strand. The corresponding propagation of the calcium transient in this
case is shown in Fig.~\ref{fig:YGapLabel}(b). Propagation block
occurs when strait width is reduced as 
illustrated in Fig.~\ref{fig:YGapComplete}(b) for $\Delta y=$0.4\,mm
and the same fibroblast count with $N=15$ as in panel (a). 
Propagation block also occurs when the fibroblast count in the
flanking fibrotic regions is increased as demonstrated in Fig.~\ref{fig:YGapComplete}(c) for a strand with $\Delta y=$0.6\,mm,
identical to that of panel (a) but with $N=18$. 
The critical threshold curve separating the regions where successful
propagation and block occur in the $(N - \Delta y)$ parameter space is
shown in Fig.~\ref{fig:YGapLabel}(a). The threshold curve tends to
$N=10$ from above  as $\Delta y$ tends to 0 mm in agreement with the
uniform case C1.

\begin{figure*}[!t]
 \includegraphics[width=\textwidth]{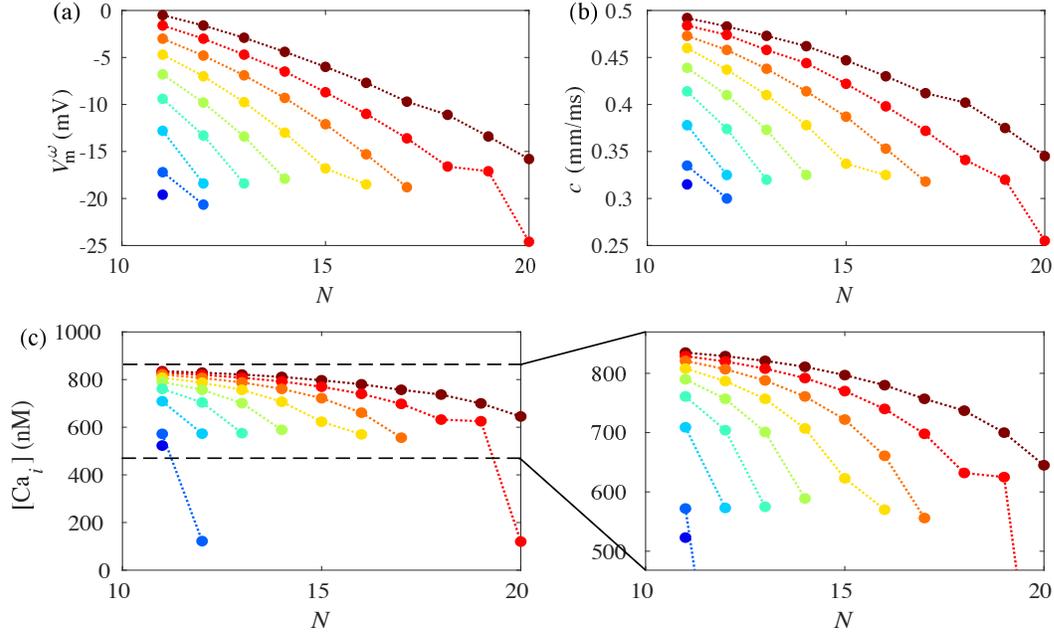}\\[3mm]
% \setlength{\tabcolsep}{0pt}
% \begin{tabular}{l@{\extracolsep{-3mm}}l}
% \begin{overpic}[width=0.52\textwidth,tics=10]{Fig09a.eps}
% \put (2,54) {{(a)}}
% \end{overpic}&
% \begin{overpic}[width=0.52\textwidth,tics=10]{Fig09b.eps}
% \put (2,54) {{(b)}}
% \end{overpic}
% \end{tabular}
% \begin{overpic}[width=1.03\textwidth,tics=10]{Fig09c.eps}
% \put (3,27) {{(c)}}
% \end{overpic}
\caption{Selected biomarkers of  myocyte action potentials propagating
  down myocyte straits \eqref{n3} as functions of the strait width
  $\Delta y$ and the number of fibroblasts $N$ coupled to each myocyte in
  the flanking regions. Dotted lines connect points
  with identical values of $\Delta y$ increasing from 0.1 mm to 0.9 mm
  in  increments of 0.1mm from bottom (blue color) to 
  top (dark red color). (a) Peak myocyte potential, (b) wave speed and
  (c) peak calcium concentration in the middle of the straits at
  $y=L/2$.
} 
	\label{fig:YGapSpeedVmax}\vspace*{-9pt}
\end{figure*}

Biomarkers of action potentials propagating down myocyte straits
are plotted in Fig.~\ref{fig:YGapSpeedVmax}.
Each line in the plots connects values with the same strand width
$\Delta y$ increasing from 0.1 mm to  0.9 mm, for a range of
fibroblast counts $N \in [11,20]$ in the flanking fibrous 
regions. As $N$ increases the peak potential, $V^\omega$, the wave speed $c$ and the peak calcium concentration
$[\text{Ca}_i]^\omega$ all decrease until the action potential is
blocked, for every value of the width $\Delta y$. Notice that 
the values of $[\text{Ca}_i]^\omega$ for $N=12$ with $\Delta y=0.2$ mm and
$N=20$ with $\Delta y=0.8$ mm are significantly smaller than adjacent
values. This is due to the proximity of these simulations to the
threshold curve shown in Fig.~\ref{fig:YGapLabel}(a).

\red{
To avoid numerical instabilities near the propagation threshold curve
the mesh discresization size was reduced from 0.1 mm to 0.05 mm in
select few cases, in particular for larger values of $N = 18,19,20$ in
Fig.~\ref{fig:YGapLabel}(a). This increase of resolution is also
responsible for an insignificant change in slope of the curves
corresponding to $\Delta y=0.8$ mm and 0.9 mm in
Fig.~\ref{fig:YGapSpeedVmax}. 
}

\section{Asymptotic theory for the case of uniform fibroblast distribution}
\label{sec:4}

In this section we extend the asymptotic theory of
\citep{2006Simitev,Biktashev2008,Simitev2011}  to the case C1 of uniform
fibroblast distribution. The theory captures qualitatively 
the behaviour of action  potential biomarkers reported above and
explains the occurrence of propagation
block with increasing fibroblast count $N$.

\subsection{Formulation of a periodic boundary value problem}
%Set-up a 1D periodic BVP
In the uniform case the fibroblast distribution takes the simple form
$n(x,y)=N$ and, by symmetry, we can neglect the dependence on the
$y$-coordinate and reduce problem \eqref{e:tissue} to one 
spatial dimension. Further, to investigate action potentials excited by a
periodic stimulus \eqref{eq:stim} and propagating with a fixed shape
and a constant speed $c$, we introduce a travelling wave ansatz $z=\tilde{x}+ct$,
where $\tilde{x}=x \sqrt{\Cm\chi/\sigma_{11}}$ is a rescaled $x$
coordinate, and arrive at the periodic boundary value problem
\begin{subequations}
\label{e:1D}
\begin{align}
c\frac{d}{dz}\Vm &= -\frac{1}{\Cm}\Big(I_{\text{m}}(\Vm,\mathbf{u})+ N\,\Gg (\Vm-\Vf)\Big) + \frac{d^2}{dz^2}\Vm,\\
c\frac{d}{dz}\Vf &= - \frac{1}{\Cf}\Big(I_{\text{f}}(\Vf,\mathbf{u}) + \Gg(\Vf-\Vm)\Big),\\
c\frac{d}{dz}\mathbf{u} &=\mathbf{F}(\mathbf{u},\Vm,\Vf),\\
\Vm(0)&=\Vm(cB), \; \frac{d}{dz}\Vm(0)= \frac{d}{dz}\Vm(cB), \;
\Vm(z_0)=\Vm^b, \;
\Vf(0)=\Vf(cB), \; \mathbf{u}(0)=\mathbf{u}(cB),
\end{align}
\end{subequations}
where $\mathbf{u}$ is a vector of all gating variables
controlling the permittivity of myocyte and fibroblast ionic channels
and of all intra- and extra-cellular concentrations of ions and
$\mathbf{F}$ are the functions controlling their dynamics as defined
in \citep{CRN98} and \citep{Morgan2016} and also specified in further
detail below. The condition
$\Vm(z_0)=\Vm^b$ is a ``phase/pinning'' condition required 
to eliminate the translational invariance of the system and arises
as a replacement of the initial value condition present in
problem \eqref{e:tissue}; here $z_0\in[0,cB]$ and $\Vm^b$ is an
arbitrary constant within range of $\Vm$, e.g.{} $\Vm^b=0$ mV.

\subsection{Asymptotic embedding}
\looseness=-1
For the analysis of the eigen-boundary value problem formulated above,
we follow an asymptotic embedding 
procedure described in \citep{Biktashev2008} and introduce in equations
\eqref{e:1D} a parameter $\epsilon > 0 $ such that for $\epsilon=1$
the embedding is identical to \eqref{e:1D} while in the limit
$\epsilon\to0^+$ useful asymptotic simplifications are obtained. There
are infinitely many ways to embed a small parameter $\epsilon$ and their merits are assessed
(a) on the basis of the usefulness of the asymptotic simplifications
and (b) on
the quality of approximation to the solutions of the original problem.
The asymptotic embedding of equations \eqref{e:1D} used below is based
on our earlier study \citep{2006Simitev} of the relative speed of the dynamical
variables in the atrial model of \cite{CRN98}. For a system of
$M$ differential equations $\d{\w_\l}/\d{\t}  = \F_\l(\w_1,\dots\w_M)$,
$\l=1,\dots M$ the relative speeds  of dynamical variables $\w_\l$ can be
formally measured by their time-scaling functions defined as
$\taul(\w_1,\ldots) \equiv \left|\d{\F_{\l}}/\d{\w_{\l}}\right|^{-1}$, 
$\l= 1\ldots M$. Comparing relevant time-scaling functions, our earlier
work established that the myocyte potential $\Vm$ and the 
gating variables $m$ and $h$ are ‘‘fast variables’’, i.e.{} they
change significantly during the upstroke of a typical action
potential, while all other variables are ``slow'' as they change only 
weakly during that period. However, an unusual
non-\cite{Tikhonov-1952} feature of the system is that $\Vm$ is both
fast 
and slow. The potential $\Vm$ is only fast because of the presence of a
large sodium  current $\gNa(\VNa-\Vm)jhm^3$. In turn, the sodium current is large
only during the upstroke of the action 
potential but not large otherwise. This is due to the near perfect
switch behaviour of gates $\m$ and $\h$ which are almost fully closed
outside the upstroke. These observations lead us to adopt the
following  asymptotic embedding of equations \eqref{e:1D}  
\begin{subequations}
\label{e:embed}
\begin{align}
\label{e:embed:a}  
c\frac{d}{dz} \Vm&= -\frac{1}{\Cm}\Big(\frac{1}{\epsilon}\gNa(\VNa-\Vm)jhm^3
+I_\Sigma(\Vm,j,\um)+ N\,\Gg (\Vm-\Vf)\Big) + \epsilon \frac{d^2}{dz^2}\Vm,\\
c\frac{d}{dz}\Vf &= - \frac{1}{\Cf}\Big(I_{\text{f}}(\Vf,\uf) + \Gg(\Vf-\Vm)\Big),\\
c\frac{d}{dz}m &=\frac{\bar{m}(\Vm,\epsilon)-m}{\epsilon\tau_m(\Vm)}, \quad \bar{m}(\Vm,0)=H(\Vm-E_m),\\
c\frac{d}{dz}h &=\frac{\bar{h}(\Vm,\epsilon)-h}{\epsilon\tau_h(\Vm)}, \quad \bar{h}(\Vm,0)=H(E_h-\Vm),\\
c\frac{d}{dz}j &=\frac{\bar{j}(\Vm)-j}{\tau_j(\Vm)},\\
c\frac{d}{dz}
\begin{bmatrix}
  \um\\
  \uf
\end{bmatrix} &=
\begin{bmatrix}
  \mathbf{T}_\text{m}(\Vm) & 0\\
  0 & \mathbf{T}_\text{f}(\Vf)
\end{bmatrix}
\left(
\begin{bmatrix}
  \overline{\um}(\Vm)\\
  \overline{\uf}(\Vf)
\end{bmatrix} -
\begin{bmatrix}
  \um\\
  \uf
\end{bmatrix} \right).
\end{align}
\end{subequations}
The current $I_\Sigma$ is the sum of all slow currents and $\um$ and $\uf$
are vectors composed of the remaining slow
gating variables (in addition to $j$ which is also slow)
with myocyte and fibroblast kinetics, respectively.
The functions $\tau_w$ and $\bar{w}$ are time-scaling
functions and quasi-stationary values of gating variables $w=j,m,h$,
respectively. For the remaining slow gates these time scaling
functions are arranged in diagonal matrices $\mathbf{T}_\text{m}$ and
$\mathbf{T}_\text{f}$ with indices denoting myocyte and fibroblast function and
$\overline{\um}$ and $\overline{\uf}$ are quasi-stationary values. The
explicit forms of these expressions are specified in \citep{CRN98,Morgan2016}. 
To account for the perfect switch behaviour of $m$ and $h$, the
functions $\bar{m}(\Vm,\epsilon)$ and $\bar{h}(\Vm,\epsilon)$ 
are ``embedded'', \ie\ they are $\eps$-dependent versions of
$\bar{m}(\Vm)$ and $\bar{h}(\Vm)$ such that $\bar{m}(\Vm;1)=\bar{m}(\Vm)$
and $\bar{h}(\Vm;1)=\bar{h}(\Vm)$ on one hand and
$\bar{m}(\Vm,0)=H(\Vm-E_m)$ and 
$\bar{h}(\Vm,0)=H(E_h-\Vm)$ on the other hand,
with $\Em=-32.7\, \mV$ and
$\Eh=-66.66\, \mV$
so that $\bar{m}(\Em)=1/2$ and $\bar{h}(\Eh)=1/2$.

\subsection{Asymptotic reduction}

% Fast-time subsystem
We are now ready to exploit the asymptotic embedding \eqref{e:embed}.
Rescaling $Z=z/\epsilon$, taking the limit $\epsilon\to0^+$ and
neglecting decoupled equations, we obtain the fast-time subsystem
\begin{subequations}
\label{e:fast}
\begin{align}
c\frac{d}{dZ}\Vm&= -\frac{\gNa \ja}{\Cm}(\VNa-\Vm)hm^3 + \frac{d^2}{dZ^2}  \Vm,\\
c\frac{d}{dZ} m &=\frac{H(\Vm-E_m)-m}{\tau_m(\Vm)},\\
c\frac{d}{dZ} h &=\frac{H(E_h-\Vm)-h}{\tau_h(\Vm)}, \\
\Vm(-\infty)&=\Va, \; \frac{d}{dZ}\Vm(\infty)= 0, \; \Vm(\infty)=\Vo, \;
\Vm(0)=\Eh, \; m(-\infty)=0, \; h(-\infty)=1.
\label{e:fastBC}
\end{align}
\end{subequations}
To avoid ambiguity we have specified in \eqref{e:fastBC} the pinning
condition explicitly at $Z=0$ and have taken $Z$ in the range 
$Z\in(-\infty,\infty)$. We have also introduced the post-front potential,
$\Vo$ as a new parameter and introduced a condition to constrain it.
We note that fibroblast kinetics does not explicitly affect this
fast-time subsystem making this very similar to the problem considered
in \citep{2006Simitev}.
% Slow-time subsystem
Taking the limit $\epsilon\to0^+$ directly in equations \eqref{e:embed}
and noting that at time scales much longer than $\epsilon$, the third
and fourth equations imply that the sodium current in equation
\eqref{e:embed:a} is proportional to $H(\Vm - E_m)H(E_h -\Vm) = 0$ which
vanishes in the limit $\epsilon\to0^+$ despite the large factor $\epsilon^{-1}$ in
front of it, we obtain the slow-time subsystem 
\begin{subequations}
\label{e:slow}
\begin{align}
c\frac{d}{dz}\Vm &= -\frac{1}{\Cm} \Big(I_\Sigma(\Vm,j,\um)+ N\,\Gg (\Vm-\Vf)\Big), \\
c\frac{d}{dz}\Vf &= -\frac{1}{\Cf} \Big(I_{\text{f}}(\Vf,\uf) + \Gg(\Vf-\Vm)\Big),\\
c\frac{d}{dz}j &=\frac{\bar{j}(\Vm)-j}{\tau_j(\Vm)},\\
c\frac{d}{dz}
\begin{bmatrix}
  \um\\
  \uf
\end{bmatrix} &=
\begin{bmatrix}
  \mathbf{T}_\text{m}(\Vm) & 0\\
  0 & \mathbf{T}_\text{f}(\Vf)
\end{bmatrix}
\left(
\begin{bmatrix}
  \overline{\um}(\Vm)\\
  \overline{\uf}(\Vf)
\end{bmatrix} -
\begin{bmatrix}
  \um\\
  \uf
\end{bmatrix} \right),\\
\Vm(0)&=\Vo, \; \Vm(cB)=\Va, \; 
\Vf(0)=\Vf(cB), \; j(0)=\ja, \; j(cB)=\ja, \nonumber \\ \um(0)&=\um(cB), \; \uf(0)=\uf(cB).
\end{align}
\end{subequations}
%
% Coupling
For a specified period of stimulation $B$, fibroblast count $N$ and
other myocyte and fibroblast parameter values, e.g.~$\Cm$, $\Cf$,
$\Gg$, $\gNa$ etc., that are all fixed as in \citep{CRN98,Morgan2016},
the coupled fast-time and slow-time systems 
\eqref{e:fast} and \eqref{e:slow} have differential equations of
cumulative order $7+\dim(\um)+\dim(\uf)$ and contain 4 free parameters ($c$, $\Va$, $\Vo$,
$\ja$), on one hand, and on the other hand, feature $10+\dim(\um)+\dim(\uf)$ boundary
conditions and 1 pinning condition. Therefore, the solution of
\eqref{e:fast} and \eqref{e:slow}, or equivalently of \eqref{e:1D}, is
fully determined and can be found using numerical methods for solution
of boundary value problems, see \citep{Simitev2011}.
 In particular, components such as $c(N)$ and $\Vo(N)$can be computed
 to compare with biomarkers from direct numerical simulations shown
 in Fig.~\ref{fig:Case1Results}. 
However, in order to understand the solutions more explicitly we next
consider the fast-time and the slow-time problems separately from each other.
\begin{figure*}[!t]
\includegraphics[width=\textwidth]{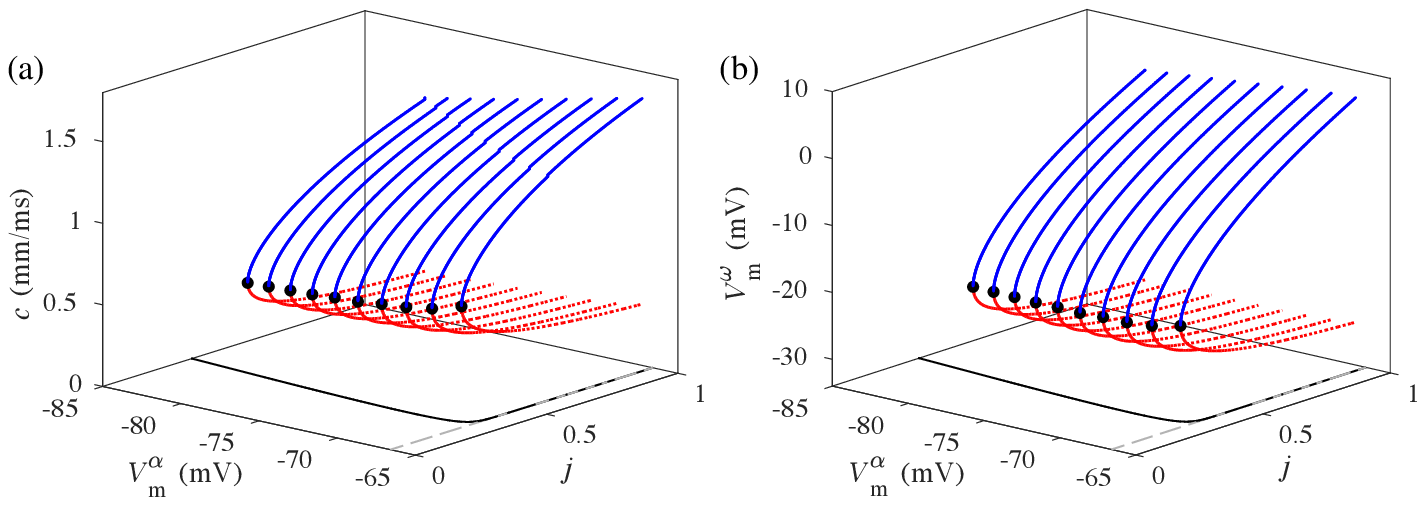}\\[5mm]
% \begin{tabular}{l@{\extracolsep{-3mm}}l}
% \begin{overpic}[width=0.52\textwidth,tics=10]{Fig10a.eps}
% \put (0,60) {(a)}
% \end{overpic}&
% \begin{overpic}[width=0.52\textwidth,tics=10]{Fig10b.eps}
% \put (0,60) {(b)}
% \end{overpic}\\
% \end{tabular}
\caption{
Selected components of the solution to the fast-time system
\eqref{e:fast} computed using the numerical method of \cite{2006Simitev}. 
(a) Wave speed $c$ and (b) peak myocyte potential $\Vo$ both as
two-valued functions of the pre-front values of the myocyte
potential $\Va$ and of the slow inactivation gating
variable of the myocyte sodium current $\ja$. 
The critical curve $\jac(\Va)$ below which the fast-time problem has
no solutions is shown as a black curve in the $(\Va-\ja)$-plane in
both panels.}
\label{fig:cjFibNCurves3D}\vspace*{-9pt}
\end{figure*}

\subsection{Solution to the fast-time system}

For fixed  myocyte and fibroblast parameter values, the fast-time
system \eqref{e:fast} has differential equations of cumulative order four
and contains four free parameters ($c$, $\Va$, $\Vo$, $\ja$), while being
constraint by five boundary conditions and one pinning condition. Therefore, the
fast-time system is expected to have a two-parameter family of
solutions, meaning that two of the free parameters can be chosen
arbitrarily and all components of the solution will be functions of
these two.
For comparison with the direct numerical simulations shown in
Fig.~\ref{fig:Case1Results}, we choose the prefront values of the
myocyte 
potential $\Va$ and of the slow inactivation gating
variable of the myocyte sodium current $\ja$ as independent parameters.
Fig.~\ref{fig:cjFibNCurves3D}	shows the wave speed and the
post-front myocyte potential as functions of the latter two,
\begin{gather}
\label{e:fastsoln}
c=c(\Va,\ja), \quad  \Vo=\Vo(\Va,\ja),
\end{gather}
respectively.
The solutions are computed using the numerical method of
\cite{2006Simitev} where a problem identical to \eqref{e:fast} save a
curvature effect term was considered. The numerics take into
account that the fast-time problem is posed on an infinite interval
and that its right-hand sides are piece-wise differentiable.
Fig.~\ref{fig:cjFibNCurves3D} shows that solutions of the
fast-time problem exist only within a certain region of the
$(\Va-\ja)$-plane above a critical curve $\jac(\Va)$. In particular, at every point within this region
two distinct solutions can be found -- one solution sitting in a
stable branch corresponding to a faster speed $c_1$ and an one
solution sitting in an unstable branch corresponding to a slower
speed $c_2$. Other solution components are similarly two-valued
functions of $\Vm$ and $\ja$. A more rigorous demonstration of these
assertions can be found in \citep{Simitev2011} where
closed-form analytical solutions are presented for a conceptual model
with a similar asymptotic structure. For the particular atrial
kinetics of \cite{CRN98} considered here, a regular perturbation
approximation of the critical curve $\jac(\Va)$ and of the wave
speed $c$ has been reported in 
\citep{2006Simitev} and a solution in terms of iterated integral expressions has been presented in \citep{Simitev2008}.

\begin{figure}[!t]
\includegraphics[width=\textwidth]{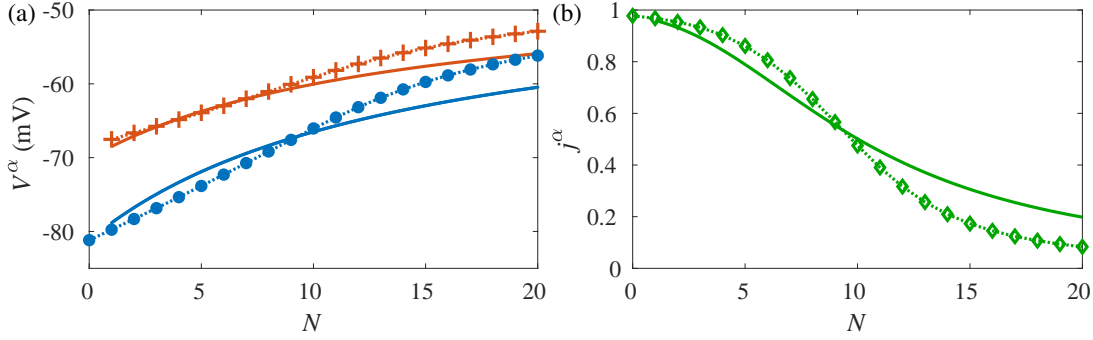}\\[3mm]
%  \begin{tabular}{l@{\extracolsep{0mm}}l}
%  \begin{overpic}[width=0.52\textwidth,tics=10]{Fig11a}
%  \put (0,57) {(a)}
%  \end{overpic}&
%  \begin{overpic}[width=0.52\textwidth,tics=10]{Fig11b}
%  \put (0,57) {(b)}
%  \end{overpic}\\
%  \end{tabular}
\caption{Selected components of the equilibrium solution to the
  slow-time system \eqref{e:slowequ}. (a)
  The resting (pre-front) myocyte potential $\Va$ (solid blue curve) and
  fibroblast potential $\Vfa$ (solid red curve) as functions of the fibroblast
  count $N$ evaluated from \eqref{e.rest}. Dotted curves with blue circle
  markers and with red plus-sign  markers show values of $\Va$ and
  $\Vfa$, respectively, measured from direct numerical simulations
  of \eqref{e:tissue} as discussed in Section \ref{UniformC1}.
  (b) The resting (pre-front) value $\ja$ of the $j$-gate as a
  function of the fibroblast count $N$ evaluated analytically (solid green
  curve) and from direct numerical simulations (dotted curve with
  diamond markers).
  Parameter values are specified in Table \ref{t:pars} with
  $B\to\infty$, and $\Gf=0.23$ nS and $\Gm=2.2$ nS. }
\label{fig:AnaVmVf1000}\vspace*{-9pt}
\end{figure}

\subsection{Equilibrium solution of the slow-time system}

For fixed  myocyte and fibroblast parameter values, the slow-time
system \eqref{e:slow} has differential equations of cumulative order
$3+\dim(\um)+\dim(\uf)$
and contains four free parameters ($c$, $\Va$, $\Vo$, $\ja$), while being constraint by
$5+\dim(\um)+\dim(\uf)$
boundary  conditions. However, wave speed
$c$ is not an essential unknown as it can be eliminated by rescaling
the independent variable and, therefore the slow-time system is
expected to have a one-parameter family of solutions. \red{A natural}
choice for the independent free parameter is the initial value of the 
myocyte potential $\Vo$, which would then, in principle, allow to determine
all slow-time solution components, for instance,
\begin{gather}
\label{e:slowsoln}
\Va=\Va(\Vo), \quad  \ja=\ja(\Vo).
\end{gather}
\looseness=-1
We note that the fibroblast kinetics is now an essential part of the
slow-time problem and the solutions also depend on other model parameters
in particular the fibroblast count $N$.
The slow myocyte and fibroblast kinetics of \citep{CRN98,Morgan2016}
are given by multi-component nonlinear expressions and analytical
solutions to problem \eqref{e:slow} are not known. To make further 
progress, we will restrict the attention to finding the 
equilibrium state of the slow-time system. This is 
sufficient to estimate wave speed and peak myocyte potential needed for
comparison to the direct numerical experiments reported in Section
\ref{UniformC1}, as well as to understand the failure of propagation
in the case of a single action potential excited by a stimulus of
infinite period $B\to\infty$, and propagating into a fully rested tissue.
\begin{figure}[!t]
\includegraphics[width=\textwidth]{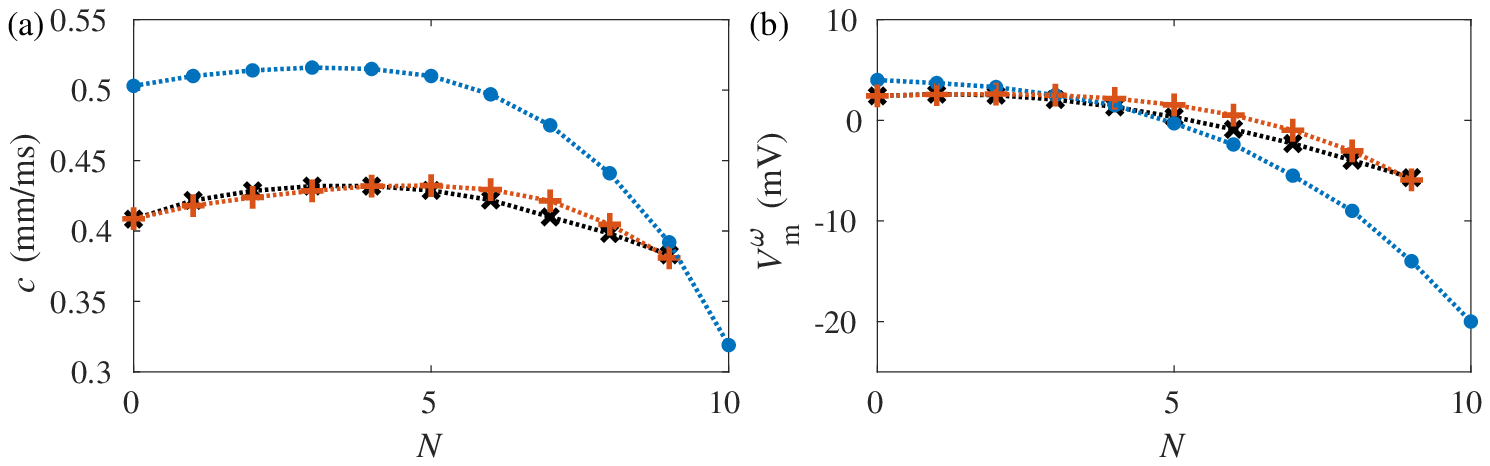}\\[3mm]
%\begin{tabular}{l@{\extracolsep{0mm}}l}
%\begin{overpic}[width=0.52\textwidth,tics=10]{Fig12a}
%\put (0,57) {{(a)}}
%\end{overpic}&
%\begin{overpic}[width=0.52\textwidth,tics=10]{Fig12b}
%\put (0,57) {{(b)}}
%\end{overpic}
%\end{tabular}
\caption{
  Selected components of the solution to the coupled fast- and
  slow-time problems. (a) Wave speed $c$ and (b) peak
  myocyte potential $\Vo$ as functions of fibroblast count $N$ for
  parameter values specified in Table \ref{t:pars}, $B\to\infty$ and $\Gf= 0.23$\,nS and $\Gm = 2.24$\,nS.
  Results   from solution of \eqref{e:fast} by the numerical method of
  \citep{2006Simitev}  coupled to expressions \eqref{e.rest} and
  \eqref{e.j0} are denoted   by black cross-sign markers.
  Results from direct numerical simulations of \eqref{e:tissue} are
  denoted with blue circle markers.
  Results   from solution of \eqref{e:fast} by the numerical method of
  \citep{2006Simitev}   coupled to $\Va$ and $\ja$ measured from
  direct numerical   simulations are denoted by red plus-sign markers.
}
	\vspace*{-9pt}
	\label{fig:FullRed}
\end{figure}

% Linearisation
The equilibrium state of the slow-time problem \eqref{e:slow} is determined by 
\begin{subequations}
\label{e:slowequ}
\begin{align}
\label{e:slowequ1}
0&= I_\Sigma(\Vm,j,\um)+ N\,\Gg (\Vm-\Vf), \\
\label{e:slowequ2}
0&= I_{\text{f}}(\Vf,\uf) + \Gg(\Vf-\Vm),\\
\label{e:slowequ3}
0&= \bar{j}(\Vm)-j,\\
\label{e:slowequ4}
0&= \overline{\um}(\Vm)-\um,\\
\label{e:slowequ5}
0&= \overline{\uf}(\Vf)-\uf.
\end{align}
\end{subequations}
Equations \eqref{e:slowequ3}, \eqref{e:slowequ4} and \eqref{e:slowequ5}
can be solved immediately,
\begin{gather}
\label{e.j0}
\ja= \bar{j}(\Vm), \quad 
\um^\alpha = \overline{\um}(\Vm), \quad
\uf^\alpha = \overline{\uf}(\Vf),
\end{gather}
and gating variables can be then eliminated from the potential equations
\eqref{e:slowequ1} and \eqref{e:slowequ2}. The latter are now involved
non-linear functions of $\Vm$ and $\Vf$ alone and linearisation near
the resting potentials $\Vm^0=-81$\,mV and $\Vf^0=-46$\,mV of the decoupled myocyte
and fibroblast models yields
\begin{subequations}
\label{e.lin}
\begin{align}
0&= I_\Sigma(\Vm)+ N\,\Gg (\Vm-\Vf) \approx \Gm(\Vm-\Vm^0)+N\Gg(\Vm-\Vf), \\
0&= I_{\text{f}}(\Vf) + \Gg(\Vf-\Vm)\approx \Gf(\Vf-\Vf^0)+\Gg(\Vf-\Vm),
\end{align}
\end{subequations}
where 
$\Gm=[d I_\Sigma/d\Vm]_{\Vm^0}\approx 2.2$ nS, $\Gf=[d
  I_{\text{f}}/d\Vf]_{\Vf^0} \approx0.23$ nS, and where we have taken
into account that the currents $I_\Sigma(\Vm^0)$ 
and $I_{\text{f}}(\Vf^0)$ vanish by the definition of $\Vm^0$ and $\Vf^0$.
Solutions to the linear set
\eqref{e.lin} are then easily obtained 
\begin{subequations}
\label{e.rest}
\begin{align}
\Va(N) & = \Vm^0 +  N (\Vf^0-\Vm^0) /\big(G_\text{m} (G_\text{gap}^{-1}+G_\text{f}^{-1}+NG_\text{m}^{-1})\big), \\
\Vf^\alpha(N) & = \Vf^0 -  (\Vf^0-\Vm^0) /\big(G_\text{f} (G_\text{gap}^{-1}+G_\text{f}^{-1}+NG_\text{m}^{-1})\big),
\end{align}
\end{subequations}
and equilibrium values of  $j$ and other gating variables are
subsequently found from
equations \eqref{e.j0}. Expressions \eqref{e.rest} and \eqref{e.j0}
are plotted in Fig.~\ref{fig:AnaVmVf1000}.
The analytical results are compared in the figure with values of
$\Va$, $\Vfa$, and $\ja$ measured from   direct numerical simulations
of \eqref{e:tissue} as discussed in   Section \ref{UniformC1} and
demonstrate accuracy adequate for the goals of this analysis.
\begin{figure}[!t]
\centering
\includegraphics[width=0.6\textwidth]{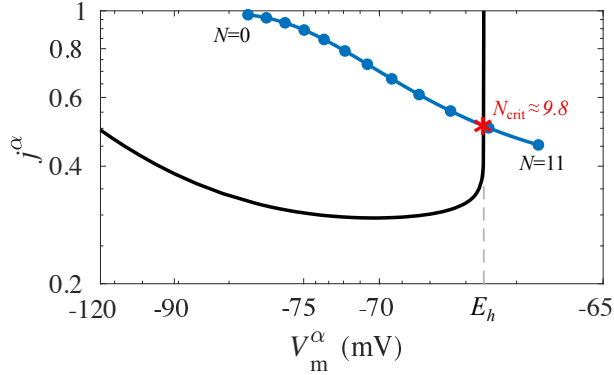}\\[3mm]
        \caption{Critical fibroblast count $\Nc$ beyond which propagation
          failure occurs. The thick solid black curve is the critical
          refractoriness boundary curve $\jac(\Va)$ for the
          existence of solutions to the fast-time system as also shown
          in Fig.~\ref{fig:cjFibNCurves3D}. The blue curve is the
          locus of pairs of equilibrium values of the
          $j$-gate and the myocyte potential 
          $\big(\Va(N),\ja(N)\big)$ for fibroblast counts $N$ as
          denoted by  circle markers along the curve. The value of $N$ at the
          intersection point marked with an asterisk is the critical
          fibroblast count $\Nc\approx=9.8$.
          Parameter values are specified in Table \ref{t:pars} with
          $B\to\infty$, and $\Gf=0.23$ nS and $\Gm=2.2$ nS.}
	  \label{fig:Ncrit}\vspace*{-9pt}
\end{figure}

\subsection{Coupling and conditions for propagation}

% Coupling 
The equilibrium solution of the slow-time system can now be coupled to
the solution of the fast-time system. The resting myocyte potential
$\Va$ and the resting value of the slow inactivation gating variable
of the myocyte sodium current $\ja$ in the tissue serve as pre-front
values for the propagating wave front. Hence, substituting 
expressions \eqref{e.rest} and \eqref{e.j0} into equations
\eqref{e:fastsoln} we find the wave speed  and the 
peak myocyte potential as functions of the fibroblast count $N$ (and
other model parameters),
\begin{gather}
c=c\big(\Va(N),\ja(N)\big),\qquad \Vo=\Vo\big(\Va(N),\ja(N)\big).
\end{gather}
In practice, expressions \eqref{e.rest} and \eqref{e.j0} are first
used to approximate $\Va(N)$ and $\ja(N)$ and these are then used as
inputs to the fast-time boundary value problem \eqref{e:fast} which is
solved by the numerical method of \citep{2006Simitev}. These asymptotic results are plotted in Fig.~\ref{fig:FullRed} and compared with the wave speed and the peak
myocyte potential measured from direct numerical simulations of
the monodomain tissue equations \eqref{e:tissue} performed as
described in section as discussed in Section \ref{UniformC1}.  
The relative error between the asymptotic approximation to the wave
speed and the values from the direct numerical simulations is
approximately 19\% at $N=0$ and decreases with increasing $N$. To
split the error contributions due to the asymptotic reduction from those due to
the linearisation of currents near the resting state, we have also plotted
in Fig.~\ref{fig:FullRed} curves computed using pre-front values
$\Va$ and $\ja$ obtained from direct numerical simulations rather than from
expressions \eqref{e.rest} and \eqref{e.j0} but still solving the
fast-time boundary value problem. Errors due to the asymptotic reduction dominate.
\begin{figure}[!t]
\includegraphics[width=1.05\textwidth]{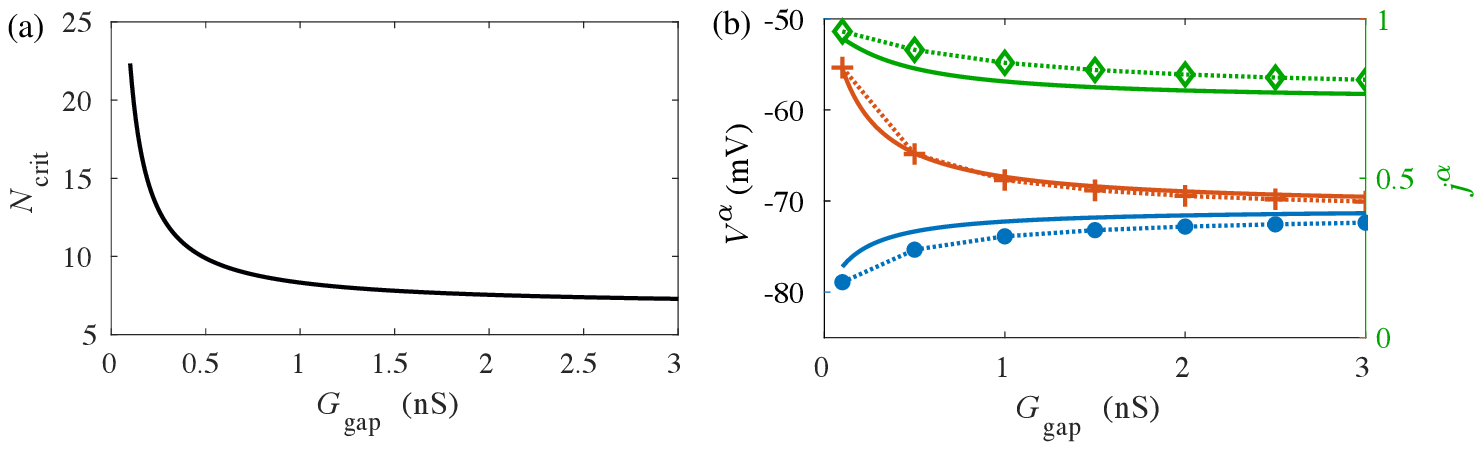}\\[3mm]
% \begin{tabular}{l@{\extracolsep{-3mm}}l}
% \begin{overpic}[width=0.52\textwidth,tics=10]{Fig14a.eps}
%  \put (0,54) {{(a)}}
% \end{overpic}&
% \raisebox{-0.5mm}{\begin{overpic}[width=0.53\textwidth,tics=10]{Fig14b.eps}
% \put (-1,54) {{(b)}}
% \end{overpic}} \\
% \end{tabular}
\caption{Dependence on myocyte-fibroblast coupling conductance
$\Gg$. (a) Minimal number of coupled fibroblasts necessary for propagation
$\Nc$ computed from equation \eqref{Ncrit}.
(b) Resting (pre-front) potentials and $j$ gate values computed from
  \eqref{e.rest} and \eqref{e.j0} as well as from direct numerical
  simulations. Same formatting conventions as in 
  Fig.~\ref{fig:AnaVmVf1000} are used with $\Va$ and $\Vfa$
   on the left $y$-axis and $\ja$ on the right $y$-axis. Fibroblast
   count is fixed to $\N=4$ in (b). In (a) and (b), other parameter
   values are as specified in Table \ref{t:pars} with $B\to\infty$ and
   $\Gf = 0.23$ nS and $\Gm =2.2$ 
  nS.} \label{fig:Ggap}\vspace*{-9pt} 
\end{figure}

% Conditions for propagation and failure
With the insight from the asymptotic reduction and consequent
coupling, the occurrence of propagation failure with increasing
fibroblast count can now be easily understood. Propagating front
solutions to the fast-time system \eqref{e:fast} exist if and only if
a point $(\Va, \ja)$, with abscissa given by the value of the pre-front myocyte
potential and ordinate given the pre-front value of the slow inactivation gating
variable of the myocyte sodium current $j$, belongs to the region in the $(\Vm,j)$-plane located above
the critical bifurcation curve $\jac(\Va)$ as plotted in
Fig.~\ref{fig:cjFibNCurves3D} while propagating front solutions do not
exist in the region under the curve. The $\jac(\Va)$ curve, therefore
serves as the boundary between absolute and relative refractoriness,
i.e. the boundary between the ability and the inability of the medium
to conduct excitation waves. Refractoriness is a fundamental characteristic of
biological excitable media, including cardiac tissues.
On the other hand, for a given fibroblast count $N$, the values of
$\Va$ and $\ja$ are determined from the solution of the slow-time
system \eqref{e:slow}. In the special case of propagation of a
single action potential through a rested tissue, these are given by
equations \eqref{e.rest} and \eqref{e.j0} for the equilibrium values
of myocyte potential and the slow inactivation gating variable of the
myocyte sodium current. Therefore, propagation is possible
if these values fall within the region of relative refractoriness
above the critical curve $\jac(\Va)$  and failure occurs if these
values fall within the region of absolute refractoriness below
$\jac(\Va)$. In Fig.~\ref{fig:Ncrit} the $\jac(\Va)$ as well
as the curve $\big(\Va(N),\ja(N)\big)$ parametrised by the fibroblast
count $N$ are plotted together in the $(\Vm,j)$ plane for the
standard set of parameter values given in Table \ref{t:pars}. The
intersection of these two curves determines the critical fibroblast
count $\Nc$ beyond which propagation failure occurs. The $\Nc$
is denoted in Fig.~\ref{fig:Ncrit} and is bracketed between $N=9$
and $N=10$ in full agreement with the results from direct numerical
simulations, see Fig.~\ref{fig:Case1Results} and discussion in
Section \ref{UniformC1}.  Fig.~\ref{fig:Ncrit} also indicates that the intersection between the two curves occurs at a point far along the vertical asymptote $\Va=\Eh$ of the fast-time system critical curve $\jac(\Va)$. 
This observation allows us to derive an explicit expression for the
critical fibroblast count as a function of the parameters of the
problem. Indeed, the algebraic equation $\Va=\Eh$, where $\Va$ is given
by \eqref{e.rest} and $\Eh=-66.6$ mV, is linear in $N$ and solving his
equation we find the approximation 
\begin{gather}
\label{Ncrit}
\Nc=\frac{\Gm\, (\Gf+\Gg)\,\redm{(\Eh-\Vm^0)}}{\Gf\Gg\,\redm{(\Vf^0-\Eh)}}.
\end{gather}

% Myocyte-fibroblast coupling conductance $\Gg$ (and other things of concern)
\looseness=-1

\red{It follows directly from equation \eqref{Ncrit} that propagation is
always possible for values of the uncoupled fibroblast resting
potential smaller than $\Eh$, i.e.~when $\Vf^0 < \Eh$. This is significant
since a range of experimental values have been reported for this
parameter, e.g.~\citep{Chilton2005}.} 
Specific biomarker values reported from the direct numerical
simulations in Section \ref{sec:3} also depend on the rest
of the model parameters. Another model parameter with value poorly 
constrained from experiments is the myocyte-fibroblast coupling
conductance $\Gg$. In the above, the value of $\Gg$ was chosen
largely for numerical convenience. 
Now, with the help of the asymptotic theory developed here the
dependence on myocyte-fibroblast coupling conductance $\Gg$ can be
constructed easily. Fig.~\ref{fig:Ggap}(a) shows the critical number
of coupled fibroblasts $\Nc$ as a function of the value of
myocyte-fibroblast  coupling conductance $\Gg$ plotted from expression
\eqref{Ncrit}. Fig.~\ref{fig:Ggap}(b) shows the variation of the
resting coupled myocyte and fibroblast potentials and the resting
value of the slow inactivation gating variable of the myocyte sodium
current $j$ with variation of $\Gg$ computed from equations
\eqref{e.rest} and \eqref{e.j0}.
Results  from direct numerical simulations of the uniformly
distributed fibroblast case at fixed $N=4$ are also plotted there and show excellent agreement.
This serves to demonstrate that the asymptotic theory remains qualitatively valid.
It has been too expensive to compute the entire
critical curve as a function of $\Gg$ by direct simulations.
We believe that the conceptual interpretation and the qualitative
conclusions of the asymptotic theory remain true for a wide range of
parameters, in the same way as for the variation with $\Gg$.

\section{Summary and conclusions}
\label{sec:5}

Cardiac fibroblasts are the most abundant type of non-myocyte cells in
the myocardium. They perform various functions, differ widely in
phenotype and are  known to be electrically active. Fibroblasts 
connect to myocytes via gap junctional channels and
evidence exists that direct electrical interaction between the two
types of cells can have arrhythmogenic effects. 
In this article we perform (a) direct numerical simulations, as well as
(b) an asymptotic analysis of action potential propagation and block, in a
model of atrial tissue with myocyte-fibroblast coupling. This is done
with the aim to understand conduction disturbances, spatially
non-uniform conduction and conduction block all of which are thought
to be key elements in the initiation and sustenance of arrhythmias.

We consider a mathematical model of fibrous atrial tissue formulated
in terms of a set of cardiac monodomain equations including a
myocyte-fibroblast coupling current. Following the work of 
\cite{Weiss2009}, we adopt an ``attachment'' approach 
to couple the human atrial myocyte model of \cite{CRN98}  to  
the mammalian fibroblast model of \cite{Morgan2016}. A key advantage
of the ``attachment'' approach is that it can be easily employed within
a homogenised continuum model such as the monodomain equations and a
variety of fibroblast distributions can be prescribed simply and
acutely. The alternative ``insertion'' approach requires that the
model is defined on a discrete grid and does not lend itself easily to
asymptotic analysis of the type report here. The atrial model of
\cite{CRN98} is chosen as partial asymptotic results were
readily available in our earlier work \citep{2006Simitev} for its
specific kinetics and because our numerical code was already validated
in this case against the benchmark  of \cite{Niederer2011}.
For direct numerical simulations of the monodomain equations we use
the Strang operator splitting method.

Using this setup we investigate three idealised fibroblast
distributions: uniform distribution, fibroblast  barrier distribution and
myocyte strait distribution, that we hypothesize are constituent
blocks of realistic fibroblast distributions. Essential action
potential biomarkers that are typically measured in
electrophysiological myocardial tissue experiments including
conduction velocity, peak  potential, action potential duration, and triangulation index are
estimated from direct numerical simulations for all idealised
distributions. Failure of action potential propagation is found to
occur at certain critical values of the parameters that define each of
the idealised  fibroblast distributions and these critical values are
accurately determined.
In the case of uniform fibroblast distribution we find that electrical
excitation fails to  propagate when 10 or more  fibroblasts are
coupled to each myocyte at the standard parameter values of our
simulations. 
As fibroblast count increases from zero, peak potential decreases while
conduction velocity slightly increases for moderate fibroblast count $N$
and then as $N$ increases further a block occurs. 
The values of APD$_{90}$ increase as the fibroblast count increases
until close to the propagation threshold when  APD$_{90}$ decreases
rapidly. Similarly, the calcium concentration stays relatively
constant as $N$ increases from 0, until $N=8$, after  which it also
falls off rapidly.  
In the case of fibroblast barrier where ``healthy" tissue is separated
by a region of increased fibroblast count our direct numerical simulations show that
propagation block is determined by both the width of the fibroblast
barrier and the count of fibroblasts coupled to each myocyte within it. 
For examples, the larger the count of coupled fibroblasts within the
region the thinner it must be to allow successful AP propagation. We proceed to
determine a threshold curve of fibroblast count versus width that
splits propagation from failure. This curve is akin to
strength-duration curves that are used elsewhere to determine the
amplitude and the duration of a stimulus current that is needed to
trigger excitation. Unlike in the first and third fibroblast
distribution cases, in this second case, \red{it is not appropriate
to} measure biomarkers as the barrier region \red{is} not big enough.
In the case of myocyte strait where channel of ``healthy" tissue
between two regions of increased fibroblast count we demonstrate that successful AP
depends on both the width of the strait and the fibroblast count in
the adjacent regions. For instance, the larger the count of coupled
fibroblasts in the surrounding regions is, the wider the strand must be to admit the
pulse across. Similarly, to the second case we construct a threshold
curve of fibroblast count versus strait width that splits propagation
from failure. When the strait width is held constant, as the
fibroblast density in the adjacent regions increases the wave speed,
peak potential and the peak calcium concentration all decrease. 

To explain these direct numerical simulation results we extend and
apply an asymptotic theory in our earlier works \citep{Simitev2011}
to the case of uniform fibroblast distribution. Action potential
biomarkers values are obtained as hybrid analytical-numerical
solutions of coupled fast-time and slow-time periodic boundary value
problems and compare well to direct numerical simulations. The 
boundary of absolute refractoriness is determined solely by the 
fast-time problem and is found to depend on the values of the myocyte
potential and the slow inactivation variable of the sodium current ahead
of the propagating front of the action potential. These quantities are
in turn estimated from the slow-time problem using a regular
perturbation expansion to find the steady state of the coupled
myocyte-fibroblast kinetics. The asymptotic theory captures with
remarkable accuracy the block of propagation in the presence of
fibroblasts.  

Our work does not consider the effect of collagen formed in the
fibrotic scar. This has been excluded here as collagen is not electrically
active and thus its main contribution is to alter the value of the
effective diffusivity and this can be accounted for by simply
rescaling the spatial variables in the governing equations. While
this is straightforward in theory, it is difficult in practise to
distinguish the effects of myocyte-fibroblast coupling we report here
from the effects due to non-uniform and anisotropic collagen
distribution.
It will be of  interest to further investigate the differences
between fibroblasts and myofibroblasts, a phenotype which has been 
linked with elevated resting potentials and larger fibroblast
capacitances  \citep{Panfilov2017}. As argued above, results will 
remain qualitatively valid but important qualitative differences may 
occur.
Our work also shows that for a large  number coupled
fibroblasts, the maximum internal myocyte calcium 
concentration is significantly less than in the fibroblast-free
case. It is calcium profile that triggers the myocyte contraction and
so, the inclusion of fibroblasts is important when modelling cardiac
muscle contraction. 
An asymptotic theory for the cases of non-uniform fibroblast distributions
$n(x,y)$ will lead to a set of spatially dependent ordinary
differential equations for the steady state of the coupled
myocyte-fibroblast model and also remains a subject for further studies.

\paragraph{Acknowledgements}
This work was supported by the UK Engineering and Physical Sciences
Research Council [grant number EP/N014642/1]. Simulations were carried
out at the UK National Supercomputing Service ARCHER.

% \setlength{\bibsep}{2.2mm}
% \bibliographystyle{apalike3}
% %\bibliographystyle{abbrvnat}
% \bibliography{refs}
% \end{document}

\setlength{\bibsep}{2.2mm}

\end{document}